\documentclass[a4paper,11pt]{article}
\usepackage{jcappub} % for details on the use of the package, please see the JINST-author-manual
\usepackage{xcolor,bm}
\usepackage{amsmath}
\usepackage{tikz}
\usepackage{comment}
\usepackage{siunitx}
\usepackage{subcaption}
\def\jcap{JCAP}

\title{Straightening the Ruler: Field-Level Inference of the BAO Scale with LEFTfield}
\def\invMpc{\,h\,{\rm Mpc}^{-1}}
\def\PSMpc{\,h^{-3}\,{\rm Mpc}^{3}}

\def\L{\Lambda}

\def\vpsi{\boldsymbol{\psi}}
\def\refeq#1{Eq.~(\ref{eq:#1})}
\def\refsec#1{Sec.~\ref{Sec:#1}}
\def\d{\delta}

\def\LEFTfield{\texttt{LEFTfield}}

\author[a]{Ivana Babić,}
\author[a]{Fabian Schmidt,}
\author[a]{Beatriz Tucci}

\affiliation[a]{Max–Planck–Institut f\"ur Astrophysik, Karl–Schwarzschild–Stra\ss e 1, 85748 Garching, Germany}

\emailAdd{i.babic@physik.uni-muenchen.de}
\emailAdd{fabians@mpa-garching.mpg.de}
\emailAdd{tucci@mpa-garching.mpg.de}

\abstract{
Current inferences of the BAO scale from galaxy clustering employ a reconstruction technique at fixed cosmology and bias parameters. Here, we present the first consistent joint Bayesian inference of the isotropic BAO scale, jointly varying the initial conditions as well as all bias coefficients, based on the EFT-based field-level forward model \LEFTfield.
We apply this analysis to mock data generated at a much higher cutoff, or resolution, resulting in a significant model mismatch between mock data and the model used in the inference.
We demonstrate that the remaining systematic bias in the BAO scale is below 2\% for all data considered and below 1\% when Eulerian bias is used for inference. Furthermore, we find that the inferred error on the BAO scale is typically 30\%, and up to 50\%, smaller compared to that from a replication of the standard post-reconstruction power-spectrum approach, using the same scales as in the field-level inference. The improvement in BAO scale precision grows towards smaller scales (higher $k$).
As a validation test, we repeat this comparison on a mock dataset that is linearly biased with respect to a 1LPT (Zel'dovich) density field, following the assumption made in standard reconstruction approaches. We find that field-level inference indeed yields the same error bar as the post-reconstruction power spectrum, which is expectd to be optimal in this case.
In summary, a field-level approach to BAO not only allows for a consistent inference of the BAO scale, but promises to achieve more precise measurements on realistic, nonlinearly biased tracers as well.
}

\begin{document}
\maketitle
\flushbottom

\section{Introduction}   

Baryon Acoustic Oscillations (BAO) represent one of the most significant observables in cosmology. Originating from the period when photons and baryons were tightly coupled, the imprints of BAO are visible in both the Cosmic Microwave Background (CMB) temperature anisotropies and the clustering of matter at later times. By measuring the apparent size of the BAO scale, $r_{s}$, in the late-time matter distribution, it is possible to estimate both the angular diameter distance and the Hubble parameter as functions of redshift. Moreover, the BAO feature imprinted in galaxy clustering is much less sensitive to the details of the galaxy-matter relation (bias) than the broad-band galaxy statistics.
However, the nonlinear evolution of the matter density field and the formation of structures does have an impact. The bulk motion of matter (and galaxies with it) leads to a damping of the BAO oscillations in the power spectrum, reducing the accuracy with which the BAO scale can be measured from the late-time clustering of galaxies, as well as inducing a small systematic shift \cite{Eisenstein_2007_modeling,Seo_2007_Fisher,padmanabhan/etal:2009,sherwin/zaldarriaga}.

Traditional BAO inference methods attempt to correct for the bulk flows
by performing a nonlinear transformation of the data, known as BAO reconstruction. 
Central to this is the estimation of displacements from the observed galaxy density field to move galaxies to their estimated initial positions \cite{Rec_Eisenstein_2007, Eisenstein_2007_rec_2, Padmanabhan_2009_rec, Baojiu_rec,
Noh_2009_reconstruction,  Tassev_2012_reconstruction, Burden_2015_reconstruction, Schmittfull_2015_reconstruction, Wang_2017_reconstruction, Schmittfull_2017_reconstruction}. This method however requires making assumptions about cosmology and bias; in a consistent inference framework one should really sample the parameters associated with these together with the BAO scale.
Here, we aim to deal with this issue by adopting a fully Bayesian field-level forward modeling approach. 
  This approach incorporates BAO reconstruction by jointly inferring the large-scale modes responsible for the bulk flow \cite{Schmidt_2019_eft}.
Crucially, it does not rely on a fiducial cosmology, as usually assumed in the reconstruction procedure. Instead, cosmological parameters are jointly and consistently inferred with the initial conditions and bias parameters.\footnote{Note that in this work, we also fix all cosmological parameters apart from the BAO scale in the inference, but all bias parameters including the linear bias $b_1$ are jointly inferred with the BAO scale.}
Furthermore, since the field-level approach does not rely on compression functions, but instead obtains the information directly from the field, it also extracts the BAO information contained in higher $n$-point functions such as the bispectrum, as opposed to the standard procedure that only uses the reconstructed power-spectrum information.

We have already discussed these issues in our previous work \cite{Babic2022} where the field-level approach for inferring the BAO scale was first introduced. The starting point is an expression for the joint posterior for the initial density field, cosmological parameters, and other model parameters, which include bias parameters and stochastic amplitudes, derived within the framework of the effective field theory (EFT) of large-scale structure (LSS) \cite{Baumann_EFT, Carrasco_EFT}.
At the heart of the field-level posterior lies the likelihood function, which is the key ingredient allowing us to access the information at the field level.
The likelihood was derived within  in a series of papers \cite{Schmidt_2019_eft, sigma8_cosmology, Cabassa_2020} and we will refer to it as the EFT likelihood or field-level likelihood. 
In every EFT theory, a fundamental component is the cutoff scale, denoted as $\Lambda$, representing the maximum wavenumber of modes considered in the calculations (in $n$-point-function-based analyses, this is often denoted with $k_{\rm max}$). While the specific value of $\Lambda$ is arbitrary, the upper limit for $\Lambda$ in the case of the EFT of LSS is the nonlinearity scale ($\Lambda_{\text{NL}} \approx 0.25 \invMpc$ at $z=0$), where perturbation theory of LSS breaks down.

The results presented here are obtained with 
the \LEFTfield\ code, a Lagrangian EFT-based forward model \cite{n-th_order_Lagrangian_forward}. Previous cosmology results with \LEFTfield\ have been presented in \cite{Schmidt_sigma8, sigma-eight-real, Kosti__2023, nguyen2024information} for the amplitude of perturbations ($\mathcal{A}_s$ or equivalently $\sigma_8$), as well as the growth rate $f$ in \cite{Stadler_2023}. 
Moreover, Ref.~\cite{beyond2pt:2024} (``Beyond-2pt challenge'') recently presented the results of a variety of inference methods applied to cosmology-blinded mock catalogs constructed by populating dark matter halos with a halo occupation distribution, where field-level inference based on \LEFTfield\ showed very competitive results for $\sigma_8$ inference.

In our previous paper on BAO \cite{Babic2022}, we inferred the BAO scale on dark matter halo samples in N-body simulations with \LEFTfield, but in the scenario of fixed initial conditions. To assess the effectiveness of the field-level approach combined with the EFT likelihood, we compared it to a power spectrum likelihood whose covariance takes into account the fixed initial conditions (i.e., no cosmic variance).
Our analysis revealed that the error bars for the BAO scale were between 1.1 and 3.3 times smaller when using the field-level likelihood as compared to the power spectrum without BAO reconstruction, where this range corresponds to the range of scales considered, $\Lambda = [0.1 - 0.25] \, h\,\text{Mpc}^{-1}$, as well as different halo samples and redshifts.

In this paper, we go a significant step further and use the field-level likelihood to perform a joint inference of the BAO scale and the initial conditions, which of course is necessary in the application to real data. Sampling initial conditions is challenging since the dimensionality of the parameter space is of the order of $10^6$. To deal with this challenge, we follow the lead of \cite{2010_FastHSampling} and use the Hamiltonian Monte Carlo (HMC) \cite{neal2011hmc} sampling method to sample the initial conditions, and slice sampler \cite{neal2000slice} to sample the cosmological parameters. We apply this analysis to mock data generated with \LEFTfield. 
 Importantly, our mock data are generated with a substantially higher cutoff than that used in the inference, as well as different bias models (with bias parameters representative of actual halo samples), so that we can test the robustness against model misspecification.

In a pioneering work, Ref.~\cite{ramanah/etal:2019} demonstrated field-level inference of cosmology and distances on mock data, obtaining very tight constraints on the expansion history. Differently to this work, however, they did not separate BAO and broad-band (Alcock-Paczy\'nski distortion) contributions. In addition, our forward model differs substantially, in that we use an EFT-based forward model, and, crucially, our results also incorporate model misspecification.

Apart from demonstrating the feasibility and robustness of BAO scale inference at the field level with \LEFTfield, the second main goal of this work is to quantify the information gain relative to the standard reconstruction algorithm \cite{Eisenstein_2007_BAO}. To this end, we adapt the reconstruction pipeline to deal with our mock data, and then perform an inference using, as closely as possible, the same cutoffs and scales as used in the field-level inference. This allows us to perform a consistent comparison of the two methods. The results of this comparison are presented in Sec. \ref{sec: compare_field_PS}.

In Section \ref{Sec: fwd_model}, we briefly introduce the forward model used in this paper. Section \ref{Sec: code} provides details of the implementation within the \texttt{LEFTfield} code. Section \ref{Sec: data} elaborates on the generation of mock data. In Section \ref{Sec: results_field}, we then present results from the joint inference of BAO along with the initial conditions,
beginning with several validation tests of the inference pipelines.
We describe the BAO reconstruction procedure, and compare field-level inference to the pre- and post-reconstruction power-spectrum analysis in Section \ref{Sec: PS_vs_Field }. The appendix contains MCMC diagnostics and further details on the field-level inference results.

\section{Forward Model}\label{Sec: fwd_model}
The idea of forward modeling is to start from the initial conditions and then model the tracer overdensity field $\delta_{g}$ at late times. The latter is defined as 
\begin{equation}
    \delta_{g}(\textbf{x}, \tau) \equiv \frac{n_g(\textbf{x}, \tau)}{\Bar{n}_g(\tau)} -1 = \delta_{g,\text{det}}(\textbf{x}, \tau) + \epsilon(\textbf{x}, \tau),
    \label{eq:bias_expansion}
\end{equation}
where $\tau$ is the conformal time, $n_g(\textbf{x}, \tau)$ is the comoving rest-frame tracer density and $\Bar{n}_g(\tau)$ is its position-independent mean. $\delta_{g,\text{det}}(\textbf{x}, \tau)$ is the deterministic part predicted by the forward model and $\epsilon(\textbf{x}, \tau)$ is the stochastic (noise) contribution.

The forward model employed here was introduced in \cite{Schmidt_2019_eft}. The main object of this model is a joint posterior for the initial density field $\delta_{\rm in}$, cosmological parameters $\theta$ and ``nuisance'' parameters (bias parameters \{$b_{O}$\} and stochastic amplitudes $\sigma$) given the data, i.e., $P(\delta_{\rm in}, \theta, \{b_{O}\}, \sigma|\delta_{g})$. There are four ingredients to this posterior:
\begin{itemize}
    \item Prior on the initial conditions
    \item Forward model for matter and gravity 
    \item Bias model
    \item Likelihood.
\end{itemize}
We will elaborate more on each of them in the following sections, but for now, let us emphasize our usage of the EFT of LSS. Relying on the EFT approach means that we have a natural occurrence of a cutoff scale $\Lambda$.
While in analytical loop calculations one typically sends this cutoff to infinity for convenience, it is necessary to keep it finite when using a field-level forward model \cite{Schmidt_sigma8}, in close analogy to lattice field theory (see \cite{2024JCAP...01..031R} for a discussion of the connection between both conventions). Thus, in our case $\Lambda$ denotes the maximum wave number included in the analysis, and $\Lambda$ is restricted to be smaller than the nonlinearity scale at which the EFT of LSS breaks down. In practice, the cutoff $\Lambda_{\text{in}}$ is applied in the initial conditions (i.e., \emph{in the free field}), as well as in the likelihood evaluation in order to only include the modes in the data that are below the cutoff $k_{\text{max}}^{\rm like}$; specifically, throughout
  this paper we keep $\Lambda_{\text{in}} = k_{\text{max}}^{\rm like} = \Lambda$.

\subsection*{Initial density field}
A single realization of the initial density field is of the form 

\begin{equation}
  \delta_{\rm in}(\boldsymbol{k}, \hat{s}) = W_{\Lambda}(k) \sqrt{P_{\text{L}}(k)} \hat{s}(\boldsymbol{k}),
  \label{eq: delta_in}
\end{equation}
where $W_{\Lambda}(k)$ is a sharp-$k$ filter which we use to ensure the proper renormalization of the evolution of large-scale modes, $P_{\rm L}$ is the linear power spectrum and $\hat{s}(\boldsymbol{k})$ is the unit Gaussian field which we use to represent the normalized initial conditions (sometimes loosely referred to as ``phases'').

\subsubsection*{Changing the size of the BAO scale}

In standard practice, the BAO method is employed to estimate the angular diameter distance to a given observed redshift.
Precisely, one  compares the predicted scale of the BAO feature to that in the data, while varying the assumed distance. As we have already discussed in \cite{Babic2022}, this type of approach cannot easily be applied to our setup, since modifying the fiducial distance would result in changes to the comoving volume of the data, and violate the periodic boundary conditions of the simulated catalog.  
Therefore, we follow a different approach, as described in \cite{Babic2022}. Instead of altering distances, we focus on rescaling the predicted comoving sound horizon. Our goal is to change the BAO size in the initial conditions and then perform forward modeling with the modified BAO scale. Since the size of the BAO scale was imprinted into the density field at early times, i.e. where linear theory applies, this is the physically correct way to modify the BAO scale. Importantly, we can keep the comoving volume and boundary conditions of the data unmodified.

To achieve this, we parametrize the linear power spectrum via
\begin{equation}\label{eq: power spectra full}
    P_{\rm L}(k, \beta) = P_{\rm L,sm}(k)[1+ A\sin (k\,\beta\,r_{\rm fid}+\phi)\exp(-k/k_{\rm D}) ],
\end{equation}
where $P_{\rm L,sm}(k)$, $A$, $\phi$, and $k_{\rm D}$ are determined by fitting Eq. \eqref{eq: power spectra full} to the linear power spectrum obtained using the CLASS code \cite{CLASS} for the fiducial cosmology. The parameter $r_{\rm fid}$ signifies the fiducial BAO scale, and $\beta$ is defined as the ratio of a proposed BAO scale size $r_s$ to its fiducial size,
\begin{equation}
    \beta = \frac{r_{s}}{r_{\rm fid}}.
\end{equation}
Changes in $\beta$ will only affect the oscillatory part of the power spectrum, while the broad-band shape will stay unaffected. Notice that for $\beta = 1$, we recover the fiducial power spectrum. 

Next, we introduce the function $f^2$ as the ratio of the linear power spectrum for the proposed values $r_s$ to the fiducial linear power spectrum,
\begin{equation}\label{eq: lin_PS_beta}
    f^{2}(k, \beta) = \frac{P_{\rm L}(k, \beta)}{P_{\rm fid}(k)} = \frac{1+ A\sin (k\,\beta\, r_{\rm fid})\exp(-k/k_{\rm D}) }{1+ A\sin (k\, r_{\rm fid})\exp(-k/k_{\rm D}) }. 
\end{equation}
This allows us to easily find the relationship between proposed and fiducial density field
\begin{equation}\label{eq: denisty}
   \delta_{\beta}(\textbf{k}, \beta) =  f(k, \beta) \delta_{\rm fid}(\textbf{k}).
\end{equation}
Therefore, the fields $\delta_{\beta}(\textbf{k}, \beta)$ and $\delta_{\rm fid}(\textbf{k})$ differ only in the size of the BAO scale; all other parameters are kept fixed to the same values.

\subsection*{Forward model for matter and gravity }\label{sec: LPT}

In this paper, we use the Lagrangian perturbation theory (LPT). If we assume that at the initial time $\tau = 0$ the particle was at the comoving position $\textbf{q}$, which we call Lagrangian position, then under the gravitational evolution it will move and we will finally observe it at the comoving Eulerian position $\textbf{x}$ at some time $\tau$. The vector field connecting those two positions is called the Lagrangian displacement field $\boldsymbol{\psi} (\textbf{q}, \tau)$ and it is the main object of LPT,
\begin{equation}
  \textbf{x} (\textbf{q}, \tau) = \textbf{q} + \boldsymbol{\psi} (\textbf{q}, \tau).
  \label{eq:disp}
\end{equation}
The evolution of the displacement field is governed by the geodesic equation describing particle trajectories in an expanding universe, coupled with the Poisson equation. To solve these equations, the components of the displacement field are treated as small parameters,
\begin{equation}
\boldsymbol{\psi}(\textbf{q}, \tau) = \sum_{n=1}^{\infty}\vpsi^{(n)}(\textbf{q}, \tau),
\end{equation}
where contributions with higher perturbative order $n$ should be successively suppressed.

One of the key advantages of LPT lies in the possibility to integrate the equations of motion for the displacement field analytically, yielding recursion relations for $\boldsymbol{\psi}^{(n)}$ as a function of $\boldsymbol{\psi}^{(m)}$ with $m < n$. This feature enables a relatively straightforward construction of the displacement field \cite{Matsubara_2008, rampf:2012, Zheligovsky_2014, rec_LPT, n-th_order_Lagrangian_forward} (see \cite{Jeong_2015} for a related construction based on Eulerian standard perturbation theory). The starting point of these recursion relations is given by
\begin{equation} \label{eq: recursion}
    \boldsymbol{\nabla}_q \cdot\boldsymbol{\psi}^{(1)}(\textbf{q},\tau)
    = - \delta^{(1)}(\textbf{q}, \tau)
    = - D_{\rm norm}(\tau) \delta_{\rm in}(\textbf{q}),
\end{equation}
where $D_{\rm norm}$ is the normalized growth factor relative to the epoch of  the linear power spectrum used in the definition of the initial conditions, \refeq{ delta_in}.
The matter density field is given fully nonlinearly by $1+\delta(x,\tau) = |\textbf{1} + \boldsymbol{\nabla}_q \boldsymbol{\psi}(q)|^{-1}$. We do not need to evaluate this equation explicitly however, since we instead obtain $\delta$ by assigning pseudo-particles to a grid at their Eulerian positions given by \refeq{disp} (see \refsec{ code}).

\subsection*{Bias Models}
To predict the biased tracer density field, we use the deterministic relation 
\begin{equation}\label{eq: Bias relation}
        \delta_{g,{\rm det}} = \sum_{O}b_{O}O,
\end{equation}
where $O$ are bias operators and $b_{O}$ their associated coefficients. At each expansion order, a finite number of operators constitute a basis. The choice of the basis is not unique and we can always alternate between different bases (provided they are complete). In this paper, we focus on the second-order expansion and utilize both Eulerian and Lagrangian bases.

\subsubsection*{Eulerian Bias}
In the case of the Eulerian bias expansion, bias operators are constructed from the matter density field obtained using LPT as described in the previous section,
\begin{equation}
    \delta_{g,{\rm det}}(\textbf{x}, \tau) = \sum_{O}b_{O}(\tau) O[\delta_{\Lambda}](\textbf{x}, \tau).
\end{equation}
Notice that an additional sharp-$k$ filter is applied to the matter density field $\delta$, putting all its modes with $k>\Lambda$ to zero, before the construction of the bias operators. In this paper, we will use the second order bias expansion including the leading higher-derivative contribution. Bias operators contributing in this case are \cite{Desjacques_2018_bias,biased_tracers_time_evolution_2015}
\begin{equation}\label{eq: Eulerian bias}
    O \in \{\delta, \delta^2, K^2, \nabla^2\delta\},
\end{equation}
where $K$ is the tidal field defined as
\begin{equation}
    K^2 \equiv (K_{ij})^2 = \left(\left[\frac{{\partial_i \partial_j}}{{\nabla^2}}  - \frac{1}{3} \delta_{ij}\right] \delta\right)^2.
\end{equation}
We also include the higher-derivative term $\nabla^2\delta$ 
which accounts for the non-locality of halo formation. The coefficients of higher-derivative operators
are  related to the spatial scale $R_{*}$ which quantifies the size of the spatial region involved in the
process of halo or galaxy formation, and its contribution is suppressed by $k^2R_{*}^2$ on large scales.

\subsubsection*{Lagrangian Bias}
In the Lagrangian bias expansion, 
\begin{equation}\label{eq: bias expansion}
        \delta^{L}_{g,{\rm det}}(\textbf{q}, \tau) = \sum_{O^{L}}b_{O^{L}}(\tau) O^{L}(\textbf{q}, \tau),
\end{equation}
the operators $O^L$ are constructed from all scalar contributions to the symmetric part of the Lagrangian distortion tensor, defined as
 \begin{equation}
    M_{ij}(\textbf{q}, \tau) \equiv \partial_{q,(i}\psi_{j)}(\textbf{q}, \tau).
\end{equation}
We do not need to include $\text{tr}[M^{(n)}]$ for $n>1$, as these terms can always be expressed in terms of scalars constructed using lower-order operators \cite{rec_LPT}. Here, we again focus solely on the second-order bias expansion. Therefore, the bias operators we use are \cite{biased_tracers_time_evolution_2015}
\begin{equation}\label{eq: Lagrangian bias}
    O^{L}\in \left\{\delta, \left(\mathrm{tr}\left[M^{(1)}\right]\right)^2, \mathrm{tr}\left[M^{(1)}M^{(1)}\right], \nabla^2\delta\right\}.
\end{equation}
Here, we have replaced $\mathrm{tr}[M^{(1)}]$ with the Eulerian density field $\d$, which allows for identification of its bias coefficient with the standard linear bias $b_{\delta}$, commonly referred to as $b_1$. 
We have once again included the leading order higher derivative operator $\nabla^2\delta$,
evaluated in Eulerian space for numerical efficiency.

\subsection*{EFT likelihood}

By construction, our model does not capture modes above the cutoff $\Lambda$ in the initial conditions. These modes introduce scatter around the mean relation $\delta_{g,\rm det}$, which is described by the stochastic contribution expressed as the field $\epsilon$ in Eq. \eqref{eq:bias_expansion}. The purpose of the field-level EFT likelihood is to absorb this  stochastic contribution.
Considering that $\epsilon$ originates from integrating out numerous independent $k$ modes, the central limit theorem ensures that this field is Gaussian at leading order. Furthermore, owing to the local nature of tracer formation, the power spectrum of $\epsilon$ is constant at leading order, with higher-derivative corrections scaling as $k^2$. These properties lead to the formulation of the
EFT likelihood
\begin{equation}\label{eq: EFT likelihood}
        \ln \mathcal{L}_{\text{EFT}}  (\delta_{g}| \hat{s}, \theta, \{b_{O}\}, \sigma) =
        -\frac{1}{2}\sum_{|\textbf{k}|\leqslant\Lambda}\Bigg[
        \ln [2\pi\sigma^{2}(k)] + 
        \frac{1}{\sigma^{2}(k)}
        |\delta_{g}(\textbf{k}) - \delta_{g, {\rm det}}[ \hat{s}, \theta, \{b_{O}\} ](\textbf{k}) |^{2}
        \Bigg].
    \end{equation}
where $\sigma^{2}$ represents the noise power spectrum, formulated in a way which ensures its positive definiteness,
    \begin{equation}\label{eq: sohastic}
       \sigma^2(k) = \sigma_{\epsilon}^2\: ( 1 + k^2\sigma_{\epsilon,2} )^2.
    \end{equation} 
    The parameter $\sigma_{\epsilon}^2$ quantifies the variance of the noise field $\epsilon$ on the discrete grid.
    %FS: a bit confusing, instead relate to P_\epsilon
    %, in the limit of $\sigma_{\epsilon,2}\to 0$. This quantity is dependent on both the grid size,  $N_g^{\Lambda}$, and the box size $L$.
    Its relationship with the noise power spectrum in the low-$k$ limit, $P_{\epsilon}$, is given by \cite{Kosti__2023}
\begin{equation}\label{eq: noise}
    P_{\epsilon} = \sigma_{\epsilon}^2 \frac{L^3}{(N_{g}^{\Lambda})^3}.
\end{equation}

The main feature to highlight in the EFT likelihood is that it compares the data $\delta_g$ and the model $\delta_{g,\rm det}$ mode by mode, or voxel by voxel, up to the maximum scale at which we trust the model. This allows the likelihood to access the full spectrum of information within the field.

\section{Code Implementation}\label{Sec: code}

In this section we briefly summarize how the forward model and likelihood are implemented in the \LEFTfield\ code. We start from the sampling of initial conditions. The initial conditions, parametrized via the field $\hat{s}$, are discretized on a grid of size $N_g^\Lambda$, and the linear density field is obtained via Eq.~\eqref{eq: delta_in}. The size of the initial grid, $N_g^{\Lambda}$, is determined by the requirement to represent all Fourier modes (for the given box size $L$) up to the cutoff $\Lambda$. Next, we construct the bias operators. This procedure slightly varies between the Eulerian and Lagrangian case. 

In the Eulerian case, we construct the LPT displacement to second order (2LPT). During the construction, all physical modes are accounted for by suitably enlarging the grid size (see \cite{fbi_tests} for a detailed discussion). We then displace a uniform grid of pseudoparticles to Eulerian space. These pseudoparticles are then assigned (see below) to obtain the Eulerian density field $\d$. This field is subsequently sharp-$k$ filtered as described above, after which the Eulerian bias operators are constructed. We resize the grid appropriately to ensure that all modes entering the likelihood are safe from aliasing.

In the Lagrangian bias case, the set of operators listed in Eq. \eqref{eq: Lagrangian bias} is constructed concurrently with the 2LPT displacement. In addition to the uniform-weight grid used to obtain the Eulerian matter density, we also assign particles with weights given by the $O_L(\bm{q})$, effectively displacing the Lagrangian operators to Eulerian space \cite{sigma-eight-real}.

The assignment schemes implemented in \LEFTfield\ are nearest-grid-point (NGP), cloud-in-cell (CIC), triangular-shaped cloud (TSC) and non-uniform-to-uniform discrete Fourier transform (NUFFT) \cite{barnett2019parallel}, which we employ as  an assignment scheme.
For the forward model in this paper, we use the NUFFT, as it converges rapidly with Eulerian grid size thanks to the kernel deconvolution. NUFFT implements the non-uniform-to-uniform discrete Fourier transform [$f(\bm x) \to \tilde f(\bm k)$]  by assigning particle positions $\bm x_i$ with weights $f(\bm x_i)$ to a supersampled grid (increasing the resolution by a factor typically in the range 1.2 to 2) using a suitable assignment kernel with compact support (of roughly 4 to 16 grid cells). It then performs an FFT on the supersampled grid, deconvolves the assignment kernel, and finally performs a grid reduction in Fourier space to yield the desired uniform discrete Fourier transform $\tilde f(\bm k)$. This method is approximate, but accuracy close to machine precision can be obtained for very reasonable computational effort. 
For the Eulerian grid size, we choose $N_g^{\rm Eul}$ such that the Nyquist frequency corresponds to $(3/2)\Lambda$, at which point the density assignment is converged to next-to-leading order \cite{fbi_tests}.

\section{Synthetic Data sets}\label{Sec: data}

The data sets used in this paper were generated using {\LEFTfield}, and denoted as Mock A and Mock B. Both data sets were generated in a box of side length $L=2000\,h^{-1}$Mpc and with a fiducial Euclidean $\Lambda$CDM cosmology characterized by $\sigma_8=0.85$, $\Omega_m=0.3$, $\Omega_\Lambda=0.7$, $h=0.7$ and $n_s=0.967$. We emphasize that both mocks contain nonlinear information, as they were generated using second-order LPT, while the distinguishing factor between them lies in the choice of the bias model. Mock A was created using the second-order Lagrangian bias, while Mock B utilized the second-order Eulerian bias. For the bias parameters, we adopted values obtained through fixed-phase inference on realizations of halo catalogues (see for example \cite{lazeyras/etal:2021}), specifically the halo mass range $\log_{10}(M/h^{-1}M_{\odot}) = 13.0-13.5$ at redshift $z=0$. The ground-truth values for the BAO scale, $\beta_{0}$, were deliberately selected to be different from 1 to verify unbiased inference even in the case of a nontrivial rescaling of the fiducial power spectrum. 
Furthermore, the mocks were generated at the cutoff $\Lambda_0 = 0.3 \, h\,\text{Mpc}^{-1}$, while the inference was performed at lower $\Lambda$ values, allowing us to test the impact of model mismatch in this dimension as well. The values of all parameters used to generate these mocks are summarized in Table~\ref{table:Mocks}.

We determine the value of $\sigma_\epsilon$ in the generation of the mock data by requiring that the \emph{inferred} noise level, at the substantially lower cutoff scales used in the inference, is roughly comparable to that of the halo sample. The inferred noise power spectrum is listed in Table~\ref{table:Mocks}. The reason for this procedure is that the noise amplitude runs very strongly with cutoff \cite{stochRG}. The actual noise power spectra employed at the scale $\L=0.3\invMpc$ in the mock generation is at the level of $[600-900] \PSMpc$.

\begin{table} \centering \begin{tabular}{|c c c c c c c c c c |} \hline \hline Mock & Bias basis & $\beta$ & $b_{\delta}$ & $b_{\nabla^2 \delta}$ & $b_{\sigma \sigma}$ & 
$b_{\text{Tr}[M^{(1)}M^{(1)}]}$ & $b_{\delta^2}$ & $b_{K^2}$ & $P_{\epsilon}^{\L=0.18 \invMpc}$\\ \hline A & Lagrangian & 0.99 & 1.21 & 0 & -0.26 & -0.18 & - & - & 1510.1 %553.83 
\\ \hline B & Eulerian & 0.98 & 1.67 & -0.07 & - & - & -0.19 & 0 & 924.44 %915.53 
\\ \hline \hline \end{tabular} \caption{Values of parameters used to produce mock datasets A and B, with the exception of $P_{\epsilon}$ for which we list the inferred value at $\L=0.18\invMpc$ in units of $\PSMpc$ (see text for a discussion). For both mocks, we use second order LPT combined with second order bias expansion and cutoff $\Lambda_0=0.3\invMpc$.} \label{table:Mocks} \end{table}

\section{Field-level BAO inference}
\label{Sec: results_field}

\subsection{Sampling method and data analysis}\label{sec: sampling}
In this section, we provide a detailed account of how we sample from the posterior defined in Section \ref{Sec: fwd_model}. We employ a combination of the Hamiltonian Monte Carlo (HMC) sampler for initial conditions, coupled with the slice sampler for the BAO scale and noise parameters in a block sampling algorithm. The bias parameters on the other hand are analytically marginalized over \cite{sigma8_cosmology}.

The initial step involves selecting a prior for the initial field $\hat{s}$. Depending on this choice, we distinguish between two scenarios,
\begin{equation}
\hat{s}(\textbf{x}) =
    \begin{cases}
        \delta_{D}^{(N_g^3)}(\hat{s} - \hat{s}_{\mathrm{true}}), & \text{FixedIC}, \\
        \mathcal{N}(0,1), & \text{FreeIC}\,,
    \end{cases}
\end{equation}
where $\hat{s}_{\rm true}$ denotes the ground-truth initial conditions, i.e. those used in the mock generation.
%In the case when $\hat{s}$ is fixed to ground truth, we are talking about FixedIC case. Alternatively, the case when the Gaussian prior for $\hat{s}$ is chosen, we denote as FreeIC.
The FixedIC case has been extensively studied in our previous work \cite{Babic2022}, while here we are focused on the FreeIC case and use the FixedIC case only as a consistency check.

The FreeIC case is challenging in particular due to the dimensionality of the parameter space of $\hat{s}$, which is of the order of $10^6$. To perform the sampling of such a large parameter space, we use the HMC sampler which is well suited for sampling in large dimensions.
The computational effort required to generate a given number of independent samples scales roughly exponentially with the dimension of the problem $N_{\text{dim}}$ in the case of classical Metropolis-Hastings MCMC; in the case of well-tuned HMC sampling, it only scales as $N^{1/4}_{\text{dim}}$ \cite{neal2011hmc}.
Because of the way HMC generates new proposals, it requires
a forward model which is differentiable with respect to the initial conditions. Fortunately, our forward model is differentiable, and the structure of the \LEFTfield\ code allows us to find its analytical derivative by performing successive applications of the chain rule.

Apart from the initial conditions, we also sample $\beta, \sigma_{\epsilon}$ and $\sigma_{\epsilon,2}$. The priors used for these parameters are
\begin{equation}  
\begin{aligned} 
\mathcal{P}(\beta) &= \mathcal{U}(0.8, 1.2),\\
\mathcal{P}(\sigma_{\epsilon}) &= \mathcal{U}(0.05, 100),\\
\mathcal{P}(\sigma_{\epsilon,2}) &= \mathcal{U}(-10^5, 10^5),
\end{aligned}
\end{equation}
where $\mathcal{U}$ denotes the uniform distribution.

When employing the likelihood marginalized over bias parameters, one MCMC chain used Gaussian priors for bias parameters. More specifically, the priors used were $\mathcal{N}(1, 5)$ for $b_{\delta}$ and $\mathcal{N}(0, 5)$ for the other bias parameters. In addition to this, we ran two other chains that employed uniform priors: $\mathcal{U}(0.01, 10)$ for $b_{\delta}$, and $\mathcal{U}(-30, 30)$ for the remaining bias parameters. We find that the bias coefficients are well constrained by the data, and the posteriors are likelihood-dominated for both chains.  Hence, despite the different priors, all chains are found to yield consistent posteriors in $\beta, \sigma_{\epsilon}$ and $\sigma_{\epsilon,2}$. Therefore, we aggregated the results from all three chains in the analysis.
For comparison, we have also performed inferences that explicitly sampled the bias parameters. The results are shown in Figs.~\ref{fig: Mock B Fourier}--{fig: Mock A Fourier} in App. \ref{sec: apendix}, and provide evidence for the above statement.

To sample cosmological parameters, we employ the univariate slice sampling technique. This involves drawing samples from the one-dimensional probability density function associated with these parameters, considering the present state of the initial conditions, $\hat{s}$. For each data set, we run a set of three chains: one starting from true initial conditions and two starting from random initial conditions. We continue running these chains until we achieve at least 100 effective samples of the $\beta$ parameter.

\begin{table}[b]
\centering
\begin{tabular}{|c  c  c  c  c |} 
 \hline
 \hline
  Mock & Forward model & $P_{\epsilon} \, [\PSMpc] $ & $\beta$ & $b_{\delta}$ \\
 \hline
C & Linear  & 271.12 & 0.98 & 1.00\\
 \hline
 D & $1^{\rm st}$ order LPT  & 271.12 & 0.98 & 1.22\\
 \hline 
 \hline
\end{tabular}
\caption{Parameter value used to produce mock datasets C and D for validation tests. Both datasets assume a cutoff $\Lambda_0=0.18\invMpc$.}\label{table:MockC}
\end{table}

\subsection{Validation tests}

To validate the reliability of both our field-level results and power spectrum-based results, we conducted a series of tests by performing both inferences on simple mock data with well-defined expected outcomes. The details and results of these tests are presented briefly here. Both tests are performed without any model mismatch, and cutoffs $\L = \L_0 = 0.18\invMpc$.

\subsubsection{Linear model}
\label{sec:linear}

We first generated a mock from a linear forward model, where the initial density field is linearly extrapolated to $z=0$, and multiplied with a single linear bias coefficient $b_\delta$.
Since the field remains linear throughout its evolution, all information about the field at $z=0$ is fully captured by the power spectrum. 
This implies that inference based solely on the power spectrum, combined with the fitting and MCMC sampling which we describe in Sec. \ref{sec: reconstruction}, should yield the same uncertainty in $\beta$  as the field-level approach.  
In other words, the error bars $\sigma(\beta)$  obtained from both inference methods should be identical. We refer to this mock as Mock C, and its details are listed in Table \ref{table:MockC}.

As can be seen from Table \ref{table:MockCD_res}, both methods give the exact same error bar, validating that the field-level analysis chain does not underestimate the error bar. 

\begin{table}[b]
\centering
\begin{tabular}{|l| c  c|} 
\hline 
 \hline
& \multicolumn{2}{c|}{Mean and 68\% CL error on $\beta$} \\
 \hline
    & Field-level inference & Power spectrum inference \\
 \hline
 Mock C (linear)  & 0.9760 $\pm$ 0.0060  & 0.9812 $\pm$ 0.0060 (pre-rec.)\ \  \\
 Mock D (1LPT)  & 0.9766 $\pm$ 0.0065  & 0.9773 $\pm$ 0.0066 (post-rec.) \\ 
\hline 
 \hline
\end{tabular}
\caption{Results of validation tests performed on Mocks C (linear) and D (1LPT). In each case, the error bars for field-level inference agrees with the power spectrum (Mock C) and post-reconstruction power spectrum (Mock D), respectively.}\label{table:MockCD_res}
\end{table}

\subsubsection{1LPT model}
\label{sec:1LPT}

The BAO reconstruction technique relies on the Zel'dovich approximation and the linear bias model to reverse the non-linear damping of the BAO feature. Consequently, if we analyze a data set generated using the Zel'dovich approximation and a linear bias model, the post-reconstruction power spectrum should recover the same information as the field-level analysis. To test this, we constructed a mock catalog, Mock D, using first-order Lagrangian perturbation theory combined with the linear bias model. The coefficients for this mock catalog are presented in Table \ref{table:MockC} as well. This comparison provides a nontrivial test of both the field-level inference analysis chain as well as our implementation of BAO reconstruction. The latter, described in detail in Sec.~\ref{Sec: PS_vs_Field }, involves some subtleties owing to how our mock data are generated.

The results of this test are presented in Table \ref{table:MockCD_res}. We see that the error bars in both cases are almost identical, indicating that the reconstruction procedure is both effective and consistent with the field-level inference, as expected in this case.

\subsection{Field-level results}
In this section, we present inference results obtained using the marginalized EFT likelihood for both Mock A and Mock B. 

\begin{figure}[t]
    \centering
        \includegraphics[width=0.6
\textwidth]{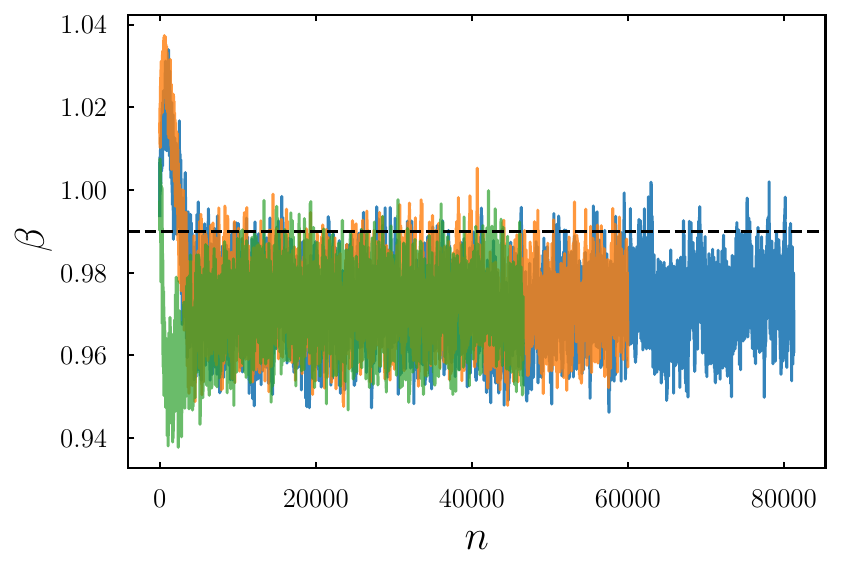}
    \caption{Trace plot of the BAO scale parameter $\beta$ for Mock A and $\Lambda = 0.2 \invMpc$ for three independent MCMC chains, where $n$ denotes the sample index. The chain shown in green started from the true initial conditions $\hat{s}_{\rm true}$ while the other two chains started from random initial conditions. Each chain started from a different initial $\beta$ value, with all quickly converging to the same value. Dashed black line indicates the ground truth value $\beta_0$. }
    \label{fig:Trace_plot_beta}
\end{figure}
\begin{figure}[t]
    \centering
        \includegraphics[width=0.8\textwidth]{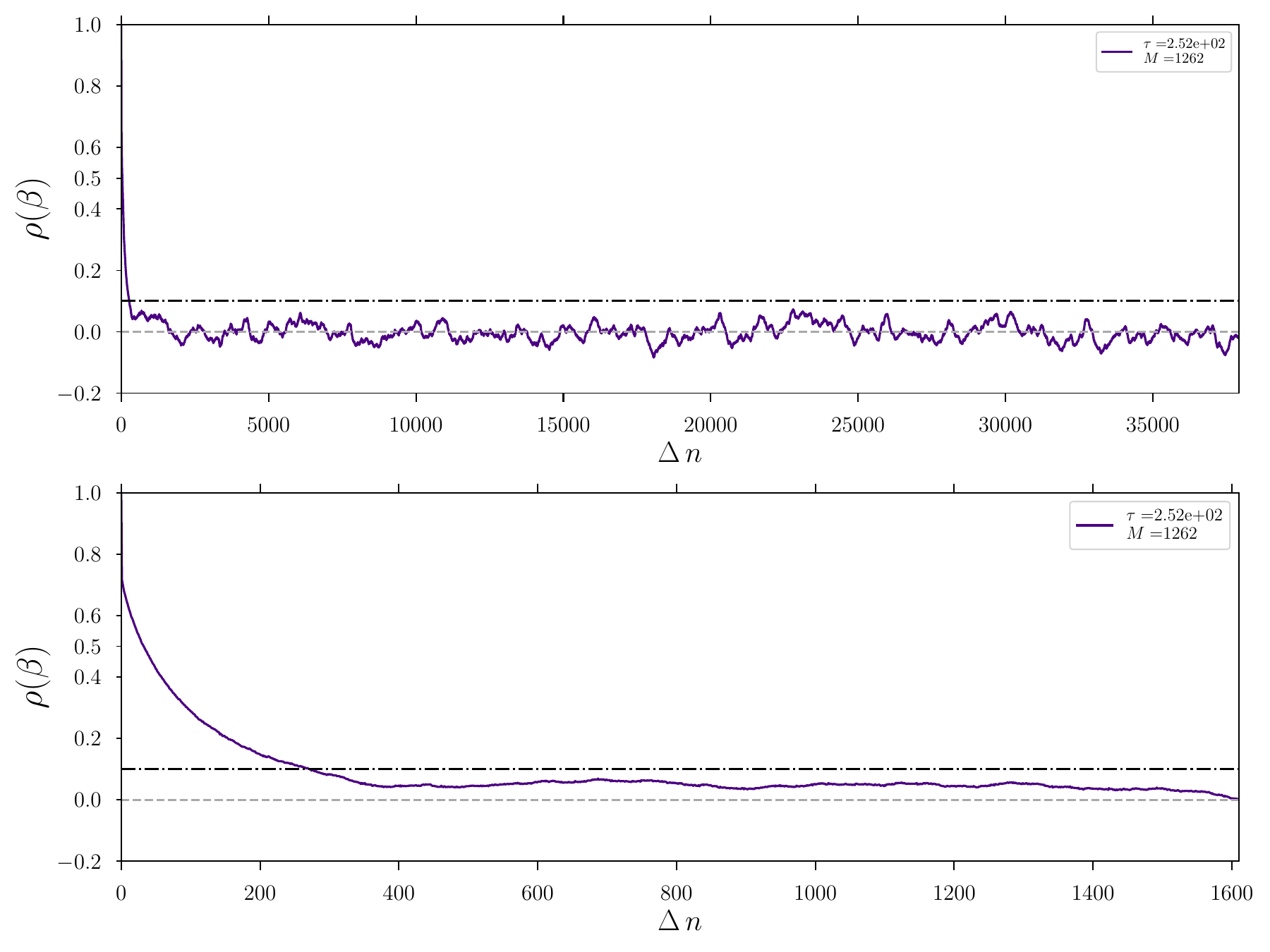}
    \caption{The normalized auto-correlation function for the BAO scale parameter $\beta$ inferred from Mock A at $\Lambda = 0.2 \invMpc$ (upper panel). We also display the correlation length value $\tau$ together with the maximum separation $M$ between the samples considered, both of which are defined in App.~\ref{sec: apendix}. The correlation length of $\beta$ is estimated to be $\tau \simeq 250$ samples. In the lower panel, we zoom into the first 1800 samples of the chain.}
    \label{fig: corr A}
\end{figure}

Mock A was generated using Lagrangian bias expansion which we also use for the inference. In Fig. \ref{fig:Trace_plot_beta}, we present a trace plot for the parameter $\beta$ corresponding to a specific value of $\Lambda = 0.2 \invMpc$. The figure shows the trajectories of all three Markov chains utilized in our analysis. Despite starting from different $\beta$ values, the chains exhibit rapid convergence towards a consistent value of $\beta$, indicating robustness in parameter estimation across varied starting points. It is also important to note that one of the chains started from the true initial conditions $\hat s_{\rm true}$ while the other two started from random $\hat s$.
Fig.~\ref{fig: corr A} depicts the normalized auto-correlation function for $\beta$ (for $\Lambda=0.2\invMpc$), for one of the chains. We find the correlation length $\tau(\beta)$ to be quite short,
substantially shorter than that of $\sigma_8$ in the field-level inference chains shown in \cite{nguyen2024information,beyond2pt:2024}. This is a major computational advantage which also allows us to push to smaller scales than used there. We determine the effective number of samples by dividing the absolute length of the chain by the estimated correlation length. 

\begin{figure}[t]
    \centering
        \includegraphics[width=1\textwidth]{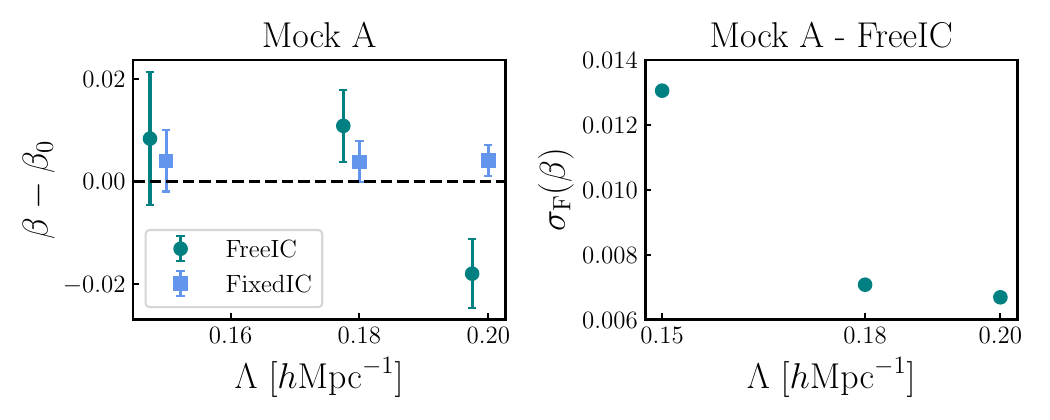}
    \caption{Field-level BAO inference results for Mock A. The left panel shows the inferred BAO scale relative to ground truth obtained using Lagranigan bias for sampling. FreeIC are represented using circle marker and FixedIC using a square. On the right we show the 68\% CL error bar on the BAO scale, $\sigma_{\rm F}(\beta)$,  as a function of cutoff $\Lambda$.}
    \label{fig: Mock_A_bias}
\end{figure}

Our main findings for Mock A are summarized in Fig.~\ref{fig: Mock_A_bias}. In the left panel, we show the inferred BAO scale, relative to the ground truth $\beta_0$. For the lower cutoffs, $\Lambda = 0.15 \invMpc $ and $\Lambda = 0.18 \invMpc$, the bias is below 1\%, however in the case of $\Lambda = 0.2 \invMpc$ the systematic offset is around 1.8\%. This is most likely due to our choice of a second-order bias expansion for the inference. The mocks were generated at a higher cutoff, which means that higher-order bias terms are effectively generated at the lower cutoff used in the inference. Moreover, higher-order bias terms are expected to become more important for higher values of $\L$. To test this assumption, we repeated the inference procedure for this mock and $\Lambda = 0.2 \invMpc$ with a third-order Lagrangian bias expansion. In this case, the remaining systematic bias in $\beta$ is reduced to less than 1\%.
%FS: update once we have sufficient samples...

The right panel of Fig. \ref{fig: Mock_A_bias} displays the 68\% confidence-level error bars for $\beta$, denoted as $\sigma_{\rm F}(\beta)$, obtained through the EFT likelihood across varying $\Lambda$ values. For the lowest cutoff, $\sigma_{\rm F}(\beta)$ slightly exceeds $1\%$, whereas for the two higher cutoffs, it reduces to $0.71\%$ and $0.66\%$, respectively.  The reduction in error bar with increasing cutoff is anticipated, as a larger $\Lambda$ permits the inclusion of more modes in both the forward model (``reconstruction'') and the likelihood (BAO inference), thereby providing additional information.
We can also compare the  behavior of $\sigma_{\rm F}(\beta)$ as a function of $\Lambda$ with the ideal expectation from mode counting, $\sigma_{\rm F}(\beta) \propto N_{\text{mode}}^{-1/2}(\Lambda)$, finding rough agreement. More precisely,  the error bar $\sigma_{\rm F}(\beta)$ improves more rapidly than the mode scaling between $\Lambda = 0.15 \invMpc$ and $\Lambda = 0.18\invMpc$, while between $\Lambda = 0.18 \invMpc$ and $\Lambda = 0.2 \invMpc$, it shrinks at a lower rate than estimated from mode counting. Note however that the simple mode counting comparison ignores the fact that $\partial P_{\rm L}(k,\beta)/\partial\beta$ also depends strongly on $k$ in an oscillatory way.

\begin{figure}[t]
    \centering
        \includegraphics[width=1\textwidth]{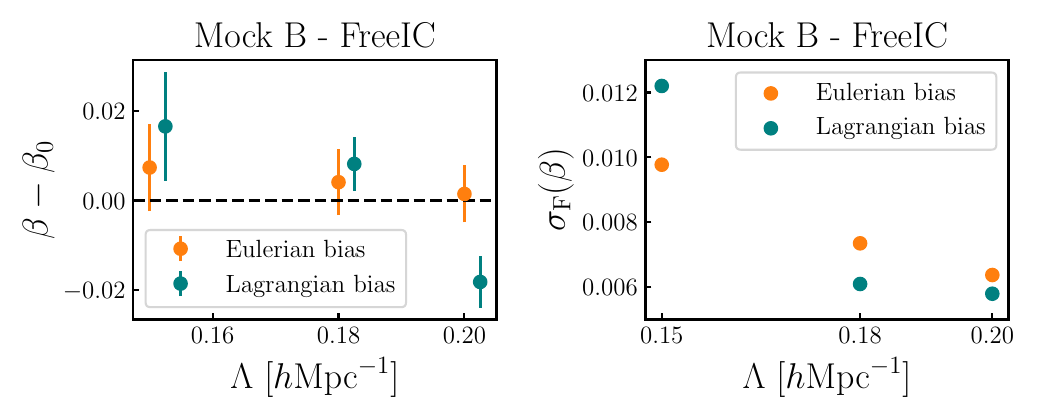}
    \caption{Field-level BAO inference results for Mock B. The left panel displays the inferred BAO scale relative to the ground truth, where the BAO scale is sampled alongside the initial conditions. The Lagrangian bias points, computed for the same values of $\Lambda$, have been slightly displaced horizontally for better visibility.
    On the right, we show the 68\% CL error bar, $\sigma_{\rm F}(\beta)$,  as a function of the cutoff $\Lambda$. The results obtained using the Lagrangian bias are depicted in blue, while those obtained using Eulerian bias are shown in orange.}
   \label{fig: Mock_B_bias}
\end{figure}

Turning our attention to Mock B, which is generated using Eulerian bias, we performed inferences with both Eulerian and Lagrangian bias. The left panel of Fig. \ref{fig: Mock_B_bias} shows the inferred $\beta$ for both Eulerian and Lagrangian bias cases. In the instance of Eulerian bias, we observe that the residual systematic bias remains below 1\% across all considered cutoffs. Additionally, the ground truth value of $\beta$ is consistently recovered within $1\sigma$.
Conversely, when using the Lagrangian bias model for inference, a slightly higher systematic bias is evident. Specifically, for $\Lambda = 0.15 \invMpc$ and $0.18\invMpc$, this bias is around 1\%, while for $\Lambda = 0.2 \invMpc$, it increases to 1.7\%. This is consistent with what we found for Mock A, and would likely be solved by going to a higher order in bias in the inference.

\begin{figure}[t]
\centering
\begin{subfigure}{.49\textwidth}
  \centering
  \includegraphics[width=\linewidth]{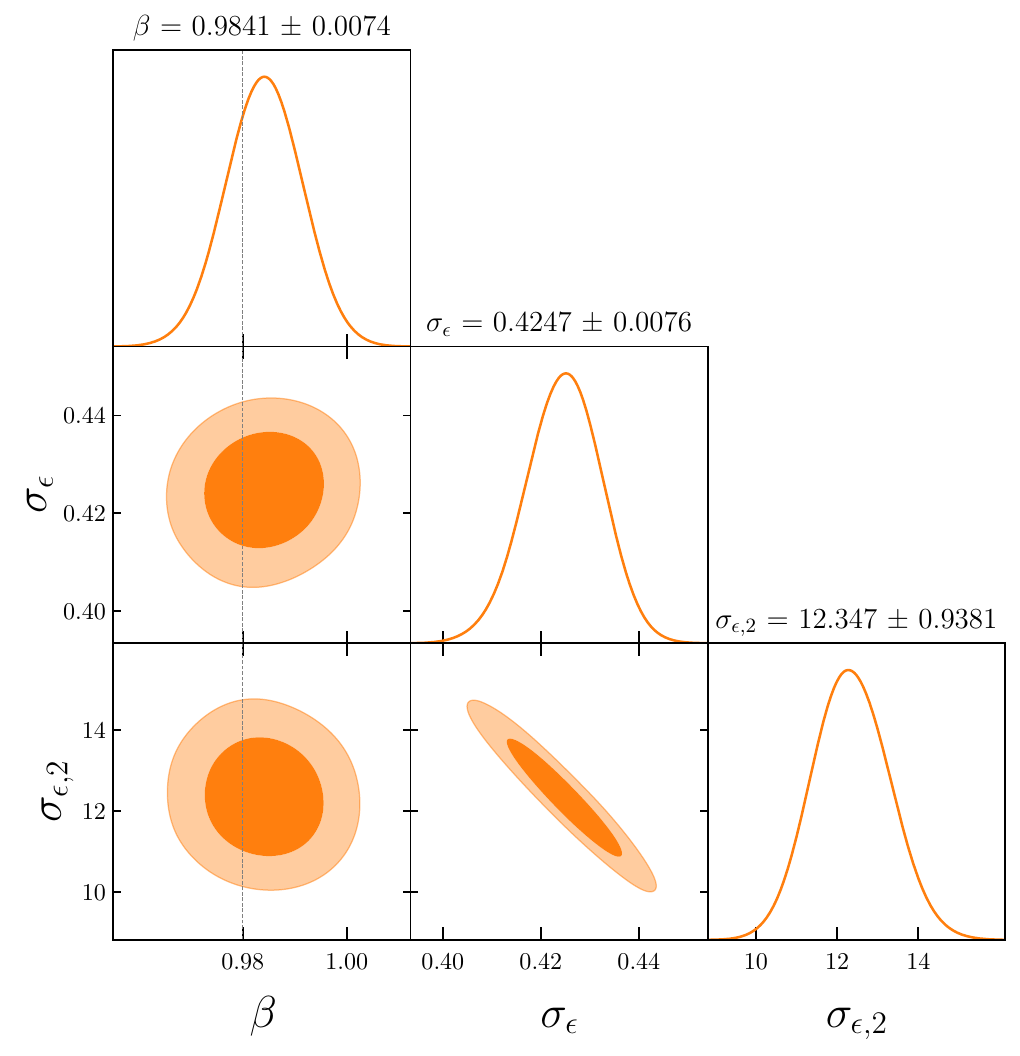}
  \caption{Eulerian bias}
\end{subfigure}%
\begin{subfigure}{.49\textwidth}
  \centering
  \includegraphics[width=\linewidth]{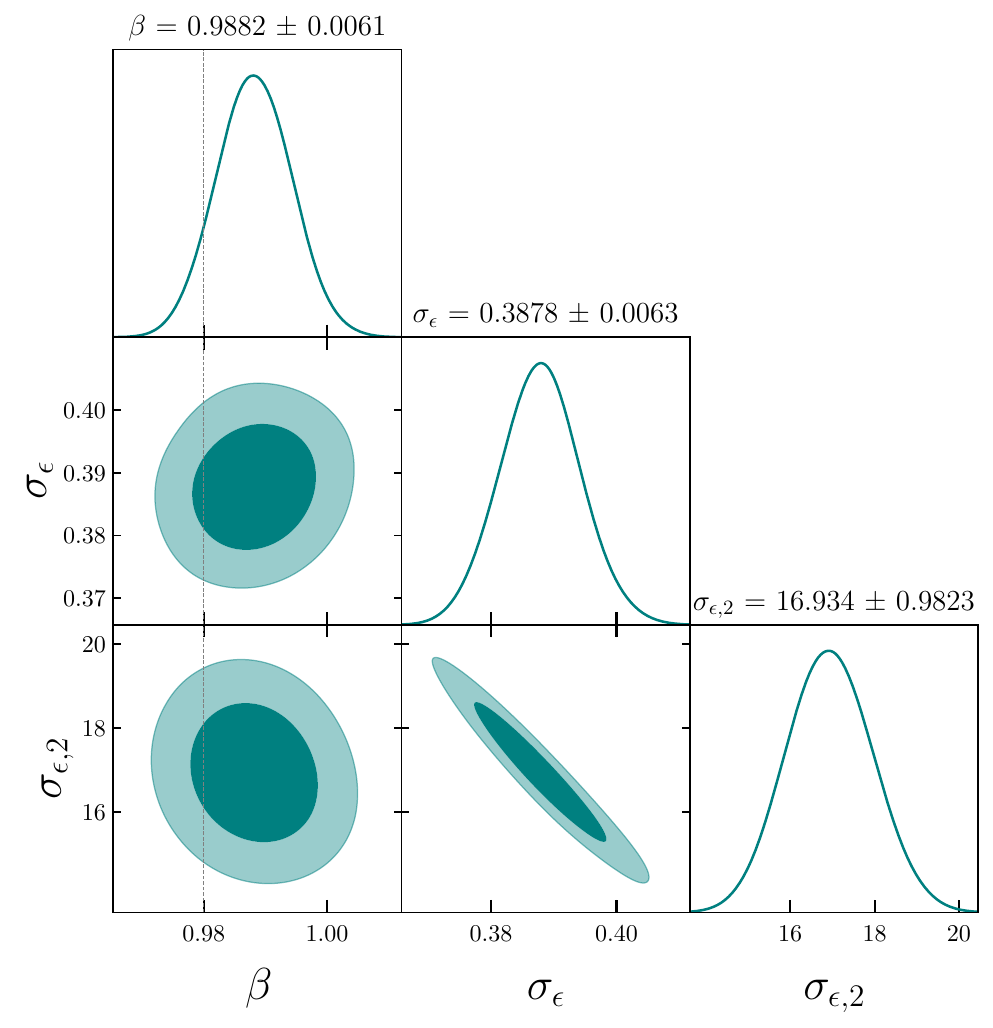}
  \caption{Lagrangian bias}
\end{subfigure}
\caption{Parameter posteriors for the FreeIC inference in the case of Mock B. Left panel (a) represents the Eulerian bias model while the right panel (b) represents the Lagrangian bias. The dotted gray line indicates the ground truth value $\beta_0$ in each case. The inference was performed at $\Lambda = 0.18 \invMpc$ in both cases.}\label{fig: corner_plot}
\end{figure}

In the right panel of Fig. \ref{fig: Mock_B_bias}, we show the inferred 68\% CL error bar $\sigma_{\rm F}(\beta)$ for Mock B. 
%We find that in the case of both bias models the size of the error bar is shrinking with increasing $\Lambda$. This trend aligns with our expectations, as a larger $\Lambda$ implies the inclusion of more modes in both the forward model and the likelihood.
Comparing the inferred error bars, $\sigma_{\rm F}(\beta)$, between the Eulerian and Lagrangian models shows that the error bar is smaller in the case of the Lagrangian bias model. This difference is likely attributable to the construction of bias operators in these models. In the Eulerian bias model, bias operators are constructed using the filtered matter field, $\delta_{\Lambda}$. This second filter (in addition to the one applied to the initial conditions) removes some mode-coupling contributions that are under control and kept in the Lagrangian bias model, which only filters the initial conditions (both bias models of course use the same likelihood filter). The downside of the Lagrangian approach is the higher computational cost due to the additional density assignments needed.

In Fig. \ref{fig: corner_plot}, we show the parameter posteriors for the Mock B  at the cutoff $\Lambda = 0.18 \invMpc$. The two panels correspond to the Eulerian and Lagrangian biases, respectively. Mock B was generated at $\Lambda_0 = 0.3 \invMpc$ with $\sigma_{\epsilon} =0.9$.
From Eq. \eqref{eq: noise}, we find that this corresponds to $\sigma_{\epsilon} \approx 0.41$ at $\Lambda = 0.18 \invMpc$.
We see that the inferred value of $\sigma_{\epsilon}$ is indeed consistent with the expected value within errors. However, we notice that $\sigma_{\epsilon,2}$ is now larger than zero, so that the effective noise level at $k\simeq \Lambda$ is increased.
This elevation in noise level is due to the fact that the inference is performed at $\Lambda < \Lambda_0$. Since the model is nonlinear, the modes between $\Lambda$ and $\Lambda_0$ which are integrated out lead to additional effective noise contributions, as shown explicitly in \cite{Kosti__2023} and at the level of renormalization-group equations in \cite{stochRG}.
%\bt{Question for Fabian: we expect the effective noise level to increase in general or specifically the non-local corrections?} \fs{Second-order bias in general increases $\sigma_{\epsilon}$ when running to lower $\L$.} FS: hopefully now more clear.

\section{Comparing field-level inference to the BAO reconstruction approach}\label{Sec: PS_vs_Field }

\subsection{BAO Reconstruction Procedure}\label{sec: reconstruction}

We now describe how we implement a BAO reconstruction pipeline on our mock data.  
We follow a standard reconstruction procedure based on the algorithm introduced by \cite{Rec_Eisenstein_2007}. However, in order for this to be applied, we need the mock data generated by the \LEFTfield\ code in the form of a catalog of discrete ``tracer'' positions. 
This catalog is constructed as follows.

We first generate a noise-free mock catalog, using the same forward model, bias parameter values, and $\beta$ values as used for the mock catalogs described in Sec.~\ref{Sec: data}.
This yields a density distribution on a grid of size $N^{\Lambda_0} = 192^3$, where $\Lambda_0 =0.3 \invMpc$ is the cutoff value used to generate the mock. 
We then Poisson sample a discrete tracer count \( N_i \) in each voxel \( i \) of the grid, where the voxel-dependent expectation value is $\langle N_i\rangle = \bar N (1 + \d_{\rm mock, noise-free})$. $\bar N = \bar n_g V_{\rm voxel}$ is determined via  the mean tracer number density $\bar{n}_g$, which we set to $\bar{n}_g = 1/P_{\epsilon}$, using the same $P_\epsilon$ as in the generation of the mocks introduced in Sec.~\ref{Sec: data}. 
Finally, we assign random positions to each of the \( N_i \) tracers within their respective voxels.

\begin{figure}[t]
    \centering
        \includegraphics[width=1\textwidth]{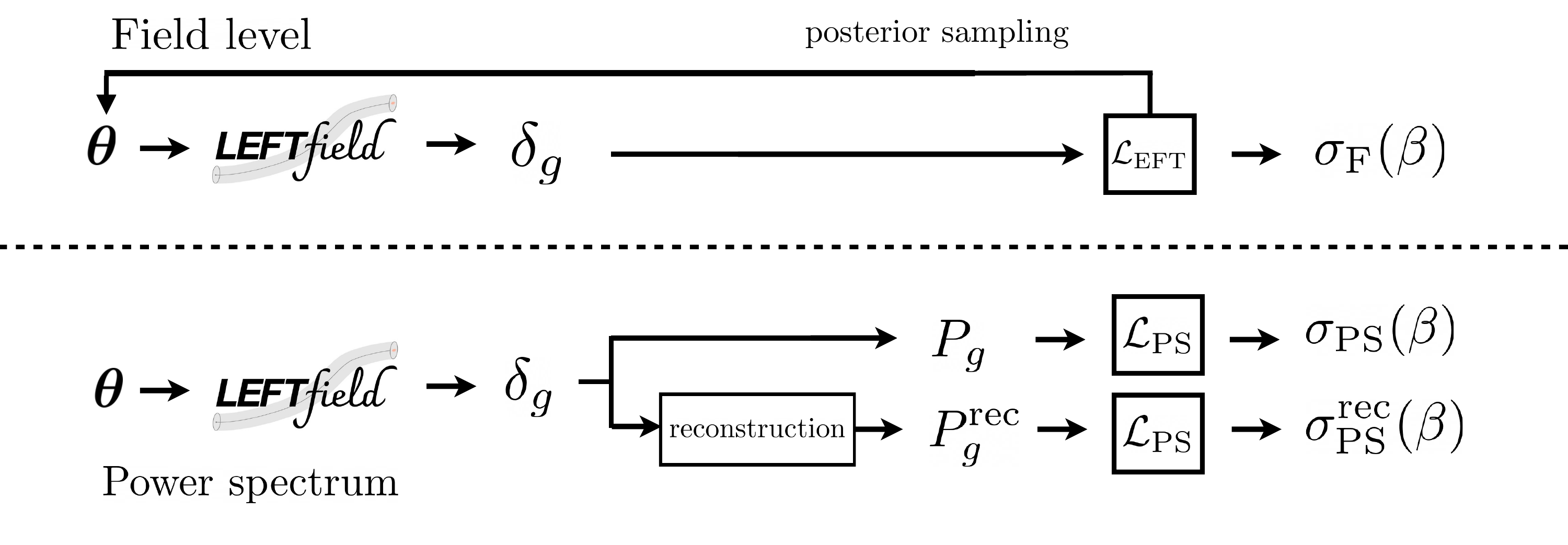}
    \caption{Flowchart of the two methods for BAO inference employed in this work: field-level inference (top) and power spectrum-based inference with and without reconstruction (bottom). The reconstruction step includes the generation of a discrete tracer catalog from $\delta_g$, as described in the text.}
    \label{fig: Recon_diagram}
\end{figure}

\begin{figure}[t]
    \centering
        \includegraphics[width=1\textwidth]{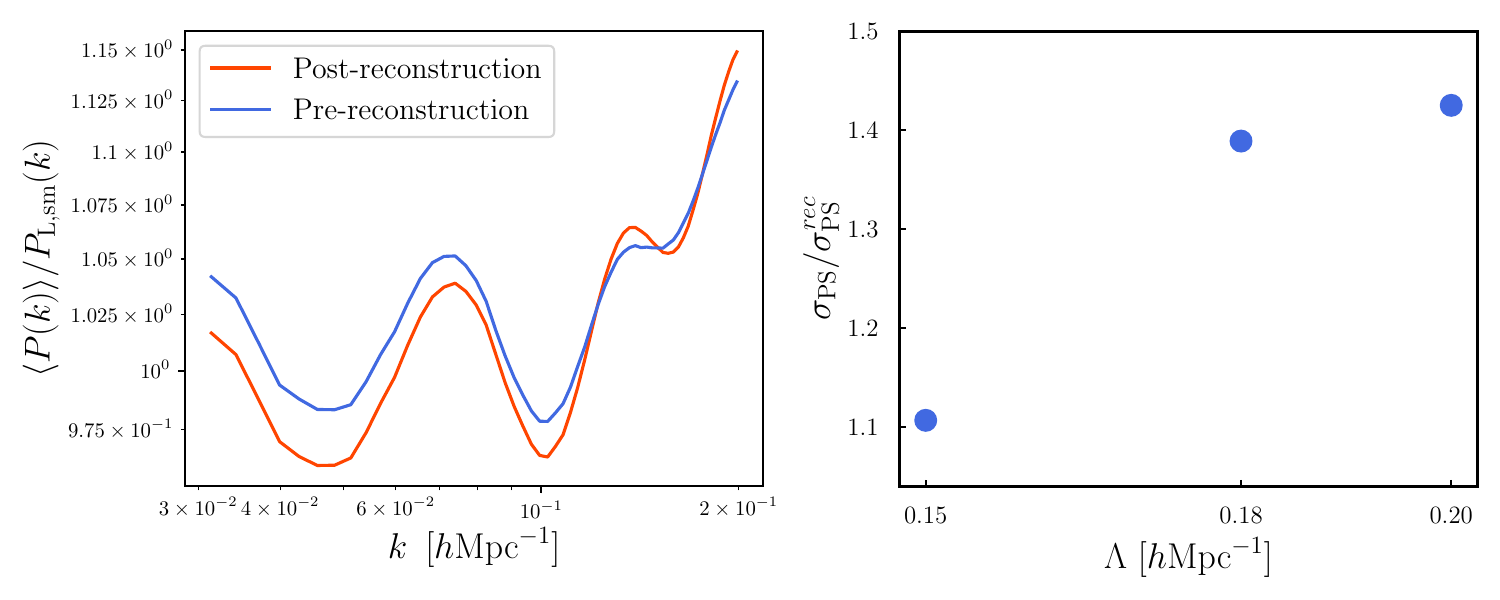}
        \caption{The left panel shows the oscillatory part of the power spectrum averaged over 1000 Mock B-like realizations. We show this for both the pre- and post-reconstructed power spectrum (for $\Lambda=0.2\invMpc$). As expected, the wiggles become more pronounced following the reconstruction. The right panel shows the inferred error bar $\sigma_\text{PS}(\beta)$ on the BAO scale using the power spectrum of the data pre-reconstruction, relative to that from the post-reconstruction power spectrum, $\sigma^{\text{rec}}_\text{PS}(\beta)$. Both inferences use the likelihood given in \refeq{L_P}.}
    \label{fig: Wiggles_post_recon}
\end{figure}

Once the mock catalog is obtained, we proceed with the reconstruction. To ensure a fair comparison with the field-level approach and guarantee that both methods have access to the same $k$-modes, we carefully select the smoothing scale and grid sizes. The choice of the smoothing scale is particularly important, given that the field-level approach employs a sharp-$k$ filter, while standard reconstruction involves a Gaussian filter.
We choose the Gaussian smoothing scale $R = \Lambda^{-1}$. We consider this to be a conservative choice for the comparison, as it allows for a significant contribution from modes with $k > \Lambda$ in the standard reconstruction approach, while these modes are excluded from the field-level analysis.
The reconstruction for a given cutoff $\L < \L_0$ proceeds as follows (see Fig. \ref{fig: Recon_diagram} for a  flowchart summarizing both analyses):
\begin{enumerate}

    \item The tracers are assigned to a grid of size $ N_{\rm assign} = 2N^{\Lambda} $ using the NUFFT assignment scheme to obtain the initial tracer field $\delta_g$.

    \item The density field $\delta_g$ is smoothed using a Gaussian filter $W_{\rm G}(k R)$ with a smoothing scale $R = 1/\L$: $\delta_g(\bm{k}) \rightarrow W_{\rm G}(k/\L)\delta_g(\bm{k})$. 

    \item Using the smoothed density field, we find the estimated displacement $\boldsymbol{\psi}$, defined as 
      \begin{equation}
        \boldsymbol{\psi}(\bm{k}) \equiv -i\frac{\bm{k}}{k^2}W_{\rm G}(k/\L)\frac{\delta_g(\bm{k})}{b_\delta},
        \label{eq:psi}
      \end{equation}
where we employ the ground-truth value for $b_\delta$ used to generate the mock.

    \item We interpolate $\boldsymbol{\psi}$ to find its value at the position of each tracer and use it to move the tracers.

    \item Once all the tracers have been shifted, we use the NUFFT assignment scheme and a grid of size $2N^{\Lambda}$ to obtain the ``displaced'' density field $\delta_d$. Notice that this operation removes a large fraction of the large-scale perturbations in $\delta_g$.
 
    \item We generate a spatially uniform grid of particles and shift them by $\boldsymbol{\psi}$ to create the ``shifted'' field $\delta_s$. The assignment scheme used to find $\delta_s$ is again NUFFT, and the grid size is $2N^{\Lambda}$.
      
    \item The reconstructed density field is obtained as $\delta_g^{\text{rec}} = \delta_d - \delta_s$, where the field $-\delta_s$ re-instates the large-scale perturbations removed from $\delta_d$.

%FS: detail that's not important for readers.      
%    \item Next, we resize the reconstructed field to a grid of size $N_{\text{g}}^{\Lambda}$ to make sure we keep only the modes below the cutoff.
      
    \item Finally, we measure the power spectrum of the reconstructed field, $P_g^{\mathrm{rec}}$, with $k_{\mathrm{max}} = \Lambda$.
\end{enumerate}
As an illustration, the left panel of Fig.~\ref{fig: Wiggles_post_recon}  displays the oscillatory part of the power spectrum averaged over 1000 Mock B-like realizations, for both pre- and post-reconstruction cases. Post-reconstruction, the wiggles become significantly more pronounced. 

We determine the BAO scale parameter $\beta$ and its uncertainty in the both pre- and post-recontruction analyses by performing an MCMC analysis using the \texttt{emcee} sampler \cite{emcee_2013}. The analysis jointly infers the BAO scale with broad-band power spectrum fitting parameters, using the following template:
  \begin{equation}
    P_{\rm model}(k, \beta) = (B_1 + B_2 k^2) P_m(k, \beta) + A(k),
    \label{eq:Pk_model}
   \end{equation}
where $B_1$ and $B_2$ are fitting parameters, and
\begin{equation}
  P_m(k, \beta) = P_{\rm fid}(k) f^2(\beta, k)
\end{equation}
represents the linear power spectrum with the rescaled BAO feature, as introduced in Eq. \eqref{eq: lin_PS_beta}.
The term $A(k)$ is modeled as a third-order polynomial:
\begin{equation}
  A(k) = a_0 + a_2 k^2 + a_3 k^3.
\end{equation}
To compute the power spectrum, we apply linear binning and perform the fit within the range $0.03 \, \invMpc < k < \Lambda$, with $\Delta k = 0.0028 \invMpc$.

The inference uses the following likelihood:
\begin{equation}
  -2\log\mathcal{L}_{\text{PS}} (\beta) = \sum_{k_i} \frac{\left(P_{\text{data}}(k_i) - P_{\text{model}}(k_i, \beta)\right)^2}{\text{Cov}[P_{\text{best-fit}}(k_i)]} + {\rm const},
  \label{eq:L_P}
\end{equation}
where the sum is over the linear bins in wavenumber with central values $k_i$, and $P_{\text{data}}(k)$ represents the power spectrum of the mock data from which we aim to infer the BAO scale.  Depending on the case considered, this can be either the pre-- or post-reconstruction power spectrum. It is worth noting that $P_{\text{model}}(k, \beta)$ already incorporates the stochastic contributions $P_\epsilon$ via the parameter $a_0$.
We use the Gaussian covariance
\begin{equation}
    \text{Cov}[P_{\text{best-fit}}(k_i)] = 2\frac{ \left[P_{\text{best-fit}}(k_i,\beta_0)\right]^2 }{m_i},
\end{equation}
where $m_i$ is the number of modes inside the bin $i$. Before the MCMC analysis , the best-fit power spectrum, $P_{\text{best-fit}}$, is obtained by applying a curve-fitting procedure to determine the best-fit parameter values for the model function given the data $P_{\rm{data}}$. We stress that these parameters are only used to determine the covariance in the likelihood \refeq{L_P}, which is then kept fixed.
Since the covariance is constant (independent of all parameters that are varied), the normalization constant in \refeq{L_P} is irrelevant.
For the parameter $\beta$, we adopt a uniform prior, $\mathcal{P}(\beta) = \mathcal{U}(0.6, 1.4)$. For the remaining parameters, we use broad Gaussian priors with means set to the values obtained in $P_{\text{best-fit}}$. We have also found that the width of these priors does not affect the $\beta$ posterior.

For each $\Lambda$ value, we run 32 chains with 6000 samples, which gives us around 2800 effective samples. 
The improvement in BAO scale determination after reconstruction is demonstrated in the right panel of Fig. \ref{fig: Wiggles_post_recon}. Notice that the relative improvement grows with with increasing $\Lambda$, which is expected since
the higher-$k$ wiggles are more affected by damping, and since the smaller
  smoothing scale employed increases the quality of the reconstruction.
%the sharpening of the BAO by reconstruction is more pronounced at higher $k$.

\subsection{Inference results and comparison with the field level}\label{sec: compare_field_PS}

We are now ready to compare the results of the power spectrum analysis (before and after reconstruction) with the field-level inference presented in the previous section.

\begin{figure}[t]
    \centering
        \includegraphics[width=0.45\textwidth]{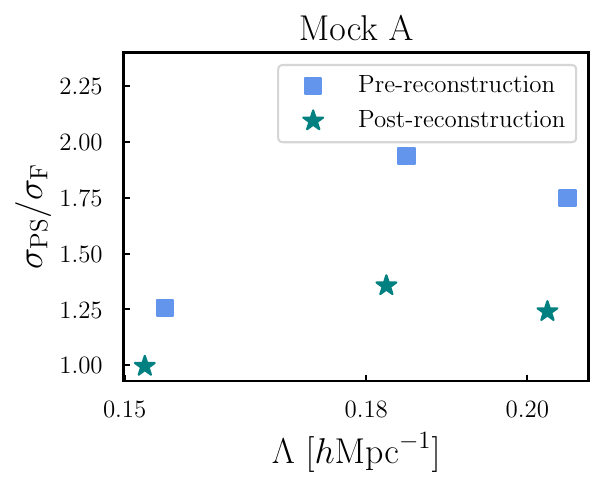}
        \caption{Inferred error bar $\sigma_\text{PS}(\beta)$ on the BAO scale using pre- and post-reconstruction power spectrum, relative to that in the field-level inference $\sigma_\text{F}$, in the case of Mock A. This mock was generated (and sampled, in case of field-level inference) using Lagrangian bias. Pre-reconstruction results are depicted using squares, whereas post-reconstruction results are represented by stars. It is evident that, even comparing to power spectrum after BAO reconstruction, the field-level BAO scale inference is more precise, by up to a factor of 1.35.}
    \label{fig: Mock A compare}
\end{figure}

We begin with Mock A, presenting in Fig. \ref{fig: Mock A compare} the ratio of error bars on $\hat{\beta}$ derived from the two power-spectrum-based analyses, $\sigma_{\text{PS}}$, to those obtained at the field level, $\sigma_{\text{F}}$.
Compared to the field level, the pre-reconstruction power spectrum yields an error bar between 1.25 and 2 times larger. As expected, this difference between $\sigma_{\text{PS}}$ and $\sigma_{\text{F}}$ decreases once we perform reconstruction.
At the lowest cutoff considered, $\Lambda = 0.15\invMpc$, field-level inference and post-reconstruction power spectrum yield identical posteriors. For the higher cutoff values, the post-reconstruction error is $1.25-1.35$ times larger than $\sigma_{\text{F}}$. That is, field-level inference improves the precision of the BAO scale by $\sim 30\%$ over standard reconstruction.
  
\begin{figure}[t]
    \centering
        \includegraphics[width=0.45\textwidth]{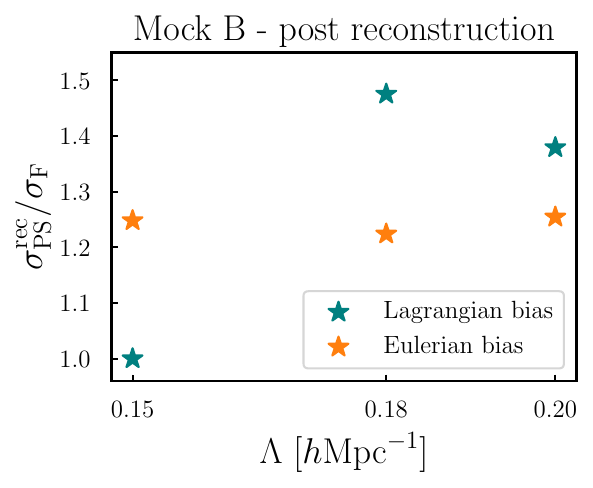}
    \caption{Inferred error bar $\sigma_\text{PS}(\beta)$ on the BAO scale from the post-reconstruction power spectrum, relative to that in the field-level inference $\sigma_\text{F}$, in the case of Mock B. Blue points indicate the scenario where Lagrangian bias is used in the field level inference, while orange points correspond to the Eulerian case (the reconstruction results are the same in both cases).}
    \label{fig: Mock B compare}
\end{figure}

We perform a similar analysis in the case of the Mock B, the results of which are summarized in Fig. \ref{fig: Mock B compare}. In the plot, the orange points represent the comparison of the post-reconstruction power spectrum variance $\sigma_{\text{PS}}$ to the case when we use the Eulerian bias expansion in the field-level inference, while blue points correspond to inferences based on the Lagrangian bias expansion (the post-reconstruction error bars are the same in both cases).
These results show that the precision of the field-level inference of the BAO scale does depend on the forward model used. In this case, Eulerian bias yields an improvement also at the lowest cutoff, while at higher cutoffs Lagrangian bias leads to a better performance (we will discuss this in the following section). 
The comparison of Fig. \ref{fig: Mock A compare} and Fig. \ref{fig: Mock B compare} also shows that, for a given forward model used in the inference, the information gain in field-level BAO inference also depends on the dataset.
Analyses of more independent mock catalogs will be necessary to definitely assess and understand both of these trends.
%\bt{Should we comment about whether/by how much the error bars increase if we use 3rd order bias expansion? Since I expect the gains of Lagrangian basis to be less signficant then.} \fs{We don't have third-order bias results with sufficiently many samples. Maybe once resubmitting to the journal...}

\section{Discussion and Conclusions}
\label{Sec:disc}

In this paper, we have presented the outcome of field-level BAO scale inferences, which jointly infer the BAO scale together with the initial conditions, bias and stochastic parameters via the \LEFTfield\ code, applied to mock data in the rest frame (without redshift-space distortions). The mock datasets were created using bias parameters taken from a fixed-initial-condition analysis on halo catalogues, and at a substantially higher cutoff (or resolution) than those used in the inference. By introducing model misspecification in this way, we attempt to make our mocks as realistic as possible. The primary distinction between the two sets of mock data is in the bias model employed for their generation: Mock A was produced using the second-order Lagrangian bias, whereas Mock B used the second-order Eulerian bias.

For Mock A, the analysis was done only using the  second-order Lagrangian  bias expansion in the inference. It showed that the systematic bias in $\beta$ stays below 1\% and is effectively negligible for all $\Lambda$ values, except for the highest cutoff, where it increases to about 1.8\%.
We have found that this bias is likely removed when moving to a third-order bias expansion in the inference. % FS: use stronger wording when chains converged.

For the Eulerian-bias Mock B, our analysis included both second-order Lagrangian and Eulerian biases. For the Eulerian bias, we find that the remaining systematic bias is consistent with zero and decreases with increasing $\Lambda$. In the case of Lagrangian bias expansion, the remaining systematic bias is slightly higher. Again, it is possible that the residual systematic shift in the Lagrangian analysis  decreases once we use a higher-order bias expansion. Furthermore, we find that the error bar in the case of the Lagrangian analysis is lower than in the case of Eulerian analysis. This is likely due to
the additional filtering employed in the construction of Eulerian bias operators. This extra filtering step results in the omission of some mode-coupling contributions, which are preserved in the Lagrangian bias model. 

The standard approach to BAO inference is based on applying a reconstruction procedure to the tracer catalog, and then measuring the BAO scale in the post-reconstruction power spectrum. In Sec. \ref{Sec: PS_vs_Field }, we detailed the application of this reconstruction algorithm to our mock data. We fixed the linear bias coefficient used in the reconstruction process, as is commonly done, thereby giving a slight benefit to the reconstruction-based analysis pipeline over field-level inference, where all bias coefficients are jointly inferred with the BAO scale. Figures \ref{fig: Mock A compare} and \ref{fig: Mock B compare} summarize the comparison between the error bars obtained from the field level analysis, $\sigma_{\text{F}}$, and those derived using the power spectrum, $\sigma_{\text{PS}}$. These figures illustrate that, depending on the mock dataset and the value of $\Lambda$, $\sigma_{\text{F}}$ is smaller than $\sigma_{\text{PS}}$ by up to a factor of two (pre-reconstruction), or up to 1.5 (post-reconstruction). Regardless of this difference, the field-level approach offers a more consistent framework for BAO inference, as it allows for the joint sampling of all parameters. Additionally, the field-level method does not require extra fitting templates such as \refeq{Pk_model} needed in the power spectrum analysis.

Our findings are further supported by the results of Sec. \ref{sec:linear} and \ref{sec:1LPT}, where we show that the error bars on $\beta$ obtained via field-level inference precisely match those from a power-spectrum-based analysis within the context of a linear forward and bias model.
  Furthermore, when performing inference on a mock generated with a 1LPT (Zel'dovich) forward model with linear bias, using the same model in the inference, the error bars from field-level inference show excellent agreement with those obtained from the post-reconstruction power spectrum, which should be optimal for this specific setup. This latter test provides a strong, nontrivial validation of both the field-level inference approach and the BAO reconstruction procedure used throughout this work.

  These results are clearly very encouraging for the field-level approach. However, one might naturally ask two questions about these findings: first, where does the additional information come from?
And second, do these results apply to actual nonlinear tracers? Part of the answer to the first question is the fact that field-level inference not only removes the large-scale displacements, but also nonlinear bias contributions in the forward evolution.
Further, our use of a 2LPT forward model means that the change of the local BAO scale in the presence of large-scale density perturbations (due to the different local expansion in the separate-universe picture) is consistently captured. On the other hand, the Zel'dovich approximation used in \refeq{psi} does not describe this effect correctly \cite{sherwin/zaldarriaga}. Both of these effects are expected to become more significant as smaller scales are included in the analysis, which could explain the increasing improvement that we find toward smaller scales (higher $\Lambda$). We will investigate the source of information gain in more detail in upcoming work.

Throughout this paper, we have attempted to make a comparison between field-level inference and BAO reconstruction on a closely matched range of scales.
It is worth noting that the maximum wavenumbers included in our analysis are somewhat lower than those chosen in current BAO reconstruction analyses. Specifically, these choose a smoothing kernel in the displacement construction on similar scales as the one used here, but then measure the post-reconstruction power spectrum up to $k_{\rm max} \sim [0.3-0.5]\, h\,{\rm Mpc}^{-1}$. This analysis assumes that the BAO are still well described by the perturbative model on scales where perturbation theory no longer applies, an assumption which needs to be carefully validated. On the other hand, the field-level approach by construction requires that the model describe all aspects of the data up to the maximum scale included, which precludes us from pushing this approach beyond perturbative scales.

To address the second question, our next step will be to transition from mock data to dark matter halos in full N-body simulations, and perform a similar comparison there.
The redshift-space modeling recently presented in \cite{Stadler_2023,stadler_2024b} will enable an anisotropic BAO scale inference in terms of line-of-sight and perpendicular components. Moreover, we plan to jointly infer the BAO scale while varying cosmological parameters governing the growth, such as $\sigma_8$ and the growth rate $f$.

Finally, the field-level forward model also enables coupling the standard BAO reconstruction approach to a consistent Bayesian inference of the BAO scale, by way of simulation-based inference (SBI) \cite{Tucci_2024}, allowing us to consistently marginalize over bias, noise and cosmological parameters. We also plan to explore this direction in the near future.

\acknowledgments
We would like to thank Andrija Kostić for his help with setting up the analysis of the MCMC chains, and Uroš Seljak for insightful discussions that led us to conduct more validation tests.
We would also like to thank Julien Lesgourgues, Ariel Sánchez, Şafak Çelik, Minh Nguyen, Ivana Nikolac and Julia Stadler for many helpful discussions.

\appendix
\section{Posterior diagnostics and detailed results}\label{sec: apendix}
\subsection{Correlation length}

In this section, we describe in detail how we conducted the analysis of field-level chains. We follow the same procedure for both mocks. For each data set, bias model, and cutoff, we ran three chains: one chain starts from the ground-truth initial conditions, while the other two chains begin from random initial conditions, with a different seed used for each to ensure they start from a distinct set of initial conditions. All three chains also have slightly different initial values for the model parameters that are being sampled. This approach ensures that the inferred value of BAO is independent of the sampler's starting point. As mentioned in Sec. \ref{sec: sampling}, one chain uses Gaussian priors for the bias parameters, while the other two uses uniform priors; however, both of these priors are essentially uninformative, as the bias parameters are well constrained by the data, as shown for the non-marginalized cases in Figs.~\ref{fig: Mock B Fourier}--{fig: Mock A Fourier}.

To perform the analysis, we combine the three independent chains into one large chain, which consists of $N$ samples; for a parameter $f$, we label the sample points as $\{f_{s}\}$,
after discarding the burn-in portion of each chain, which is about 5 correlation lengths.
To denote the mean of the dataset, we use $\langle f_s \rangle \equiv \bar{f} = \frac{1}{N}\sum_{s=1}^N f_s$. 
We continue running the chains until we achieve at least 100 effective samples of the $\beta$ parameter for each dataset and $\Lambda$ considered. The number of effective samples is calculated by dividing the total number of samples, $N$, by the auto-correlation length. The auto-correlation length for an MCMC chain indicates the number of steps required within the chain to obtain samples that are independent of each other. Correlation length can be estimated using the relation 
\begin{equation}
    \hat{\tau}_f (M) = 1 + 2\sum_{\tau = 1}^{M} \hat{\rho}_f (\tau),
\end{equation}
where $\hat{\rho}$ is the normalized autocorrelation function
\begin{equation}
    \rho(\tau) = \frac{\mathcal{A}(\tau)}{\mathcal{A}(0)},
\end{equation}
and 
\begin{equation}
    \mathcal{A}(\tau) = \langle
f_s f_{s+\tau} \rangle_s - \langle f_s \rangle_s^2.
\end{equation}
is the auto-correlation function. $M$ is the maximum separation between the samples considered. As the best choice for $M$, Ref.~\cite{Sokal1996MonteCM} suggests using the smallest value of $M$ such that $M \geqslant C\hat{\tau}_f (M)$ for a constant $C$ chosen to minimize the covariance of the estimator. This is usually achieved for values of $C$ close to 5. Plots indicating $\tau_f$ values of $\beta$ for chains analyzed in the paper are shown in Fig. \ref{fig: corr A} and Fig. \ref{fig: corr B}. 
In the tables below, we summarize the correlation lengths for the parameter $\beta$.
There is significant scatter among the estimates for $\tau$, but all are significantly below 1,000.

\begin{table}[ht] \centering \begin{tabular}{|l l l l|} 
\hline  $\Lambda \,[\invMpc]$ & $0.15$ & $0.18$ & $ 0.2$ \\ 
\hline   &  & $\tau$ &  \\
\hline Chain 1 & 210 & 683 & 101 \\ 
\hline Chain 2 & 165 & 728 & 132 \\ 
\hline Chain 3 & 175 & 609 & 261 \\ 
\hline \end{tabular} \caption{Correlation length values for chains run on Mock A using Lagrangian bias.} \label{T: T1} \end{table}

\begin{table}[ht] \centering \begin{tabular}{|l l l l|} 
\hline  $\Lambda \,[\invMpc]$ & $0.15$ & $0.18$ & $ 0.2$ \\ 
\hline   &  & $\tau$ &  \\
\hline Chain 1 & 350 & 252 & 507  \\ 
\hline Chain 2 & 272 & 771 & 728 \\ 
\hline Chain 3 & 212  & 123 & 609 \\ 
\hline \end{tabular} \caption{Correlation length values for chains run on Mock B using Lagrangian bias.} \label{T: T2} \end{table}

\begin{table}[ht] \centering \begin{tabular}{|l l l l|} 
\hline  $\Lambda \,[\invMpc]$ & $0.15$ & $0.18$ & $ 0.2$ \\ 
\hline   &  & $\tau$ &  \\
\hline Chain 1 & 186 & 590 & 735  \\ 
\hline Chain 2 & 105 & 606 & 744 \\ 
\hline Chain 3 & 800  & 522 & 710 \\ 
\hline \end{tabular} \caption{Correlation length values for chains run on Mock B using Eulerian bias.} \label{T: T3} \end{table}

\subsection{Parameter posteriors: marginalized likelihood}
\begin{figure}[h]
  \begin{subfigure}[b]{0.49\textwidth}
    \centering
    \includegraphics[width=\linewidth]{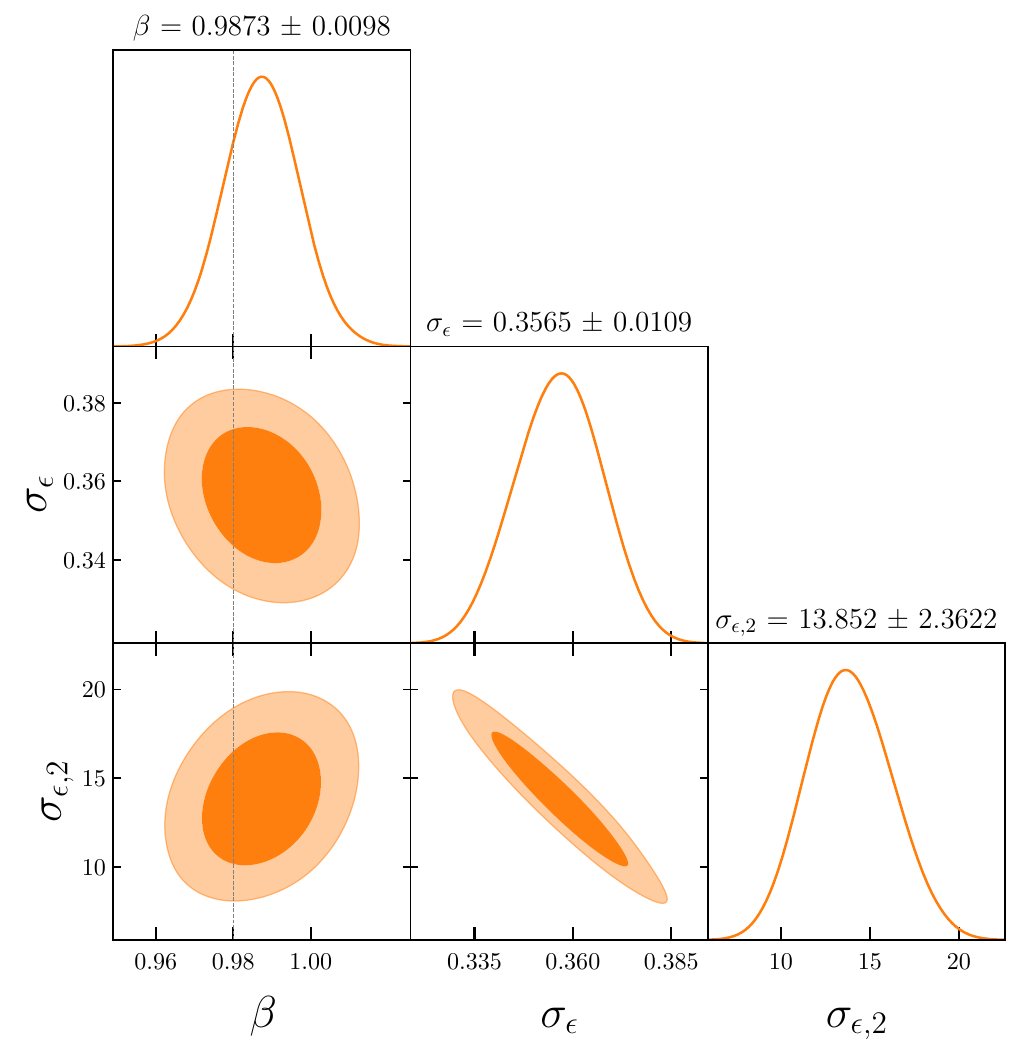} 
    \caption{Eulerian bias, $\Lambda = 0.15 \invMpc$.} \label{fig: marg_res}
    \vspace{4ex}
  \end{subfigure}%% 
  \begin{subfigure}[b]{0.49\textwidth}
    \centering
    \includegraphics[width=\linewidth]{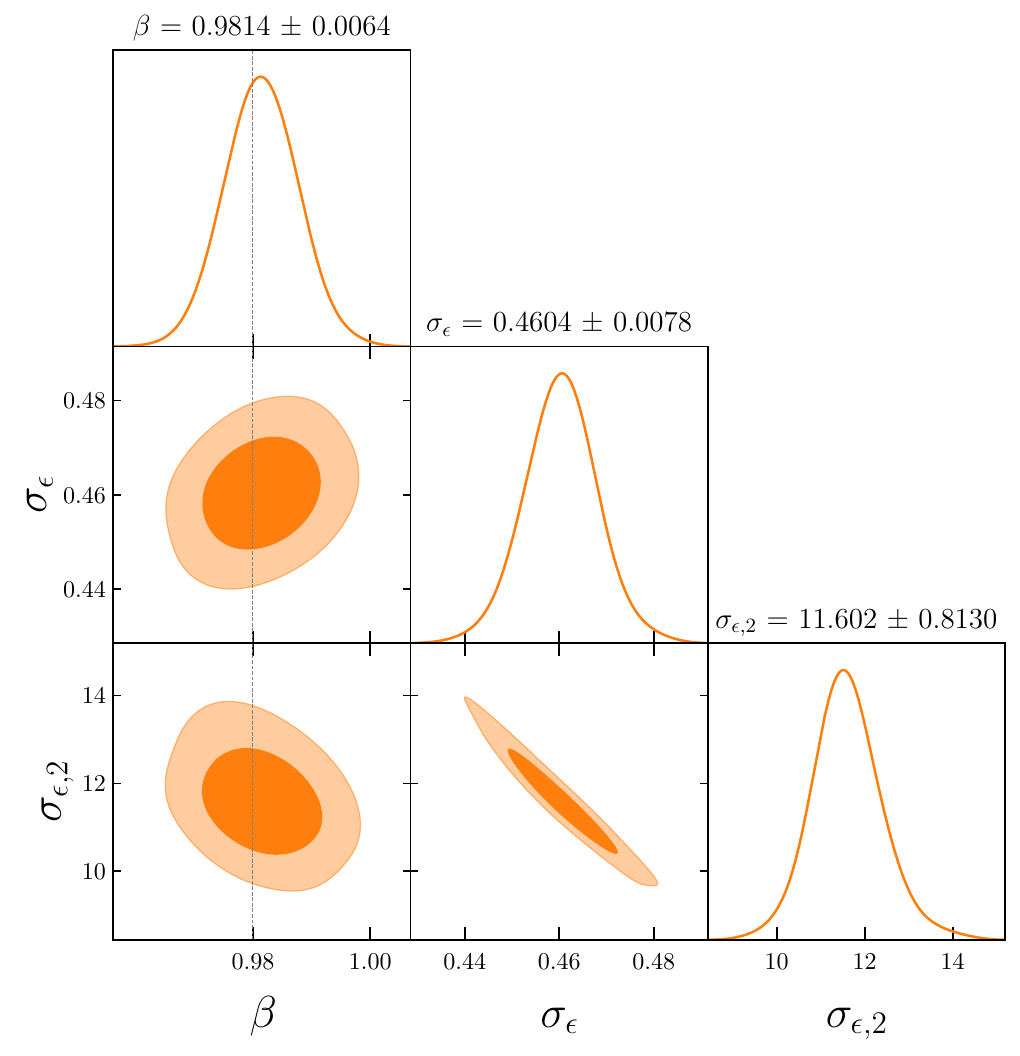 }
    \caption{Eulerian bias, $\Lambda = 0.2 \invMpc$.}  
    \vspace{4ex}
  \end{subfigure} 
  \begin{subfigure}[b]{0.49\textwidth}
    \centering
   \includegraphics[width=\linewidth]{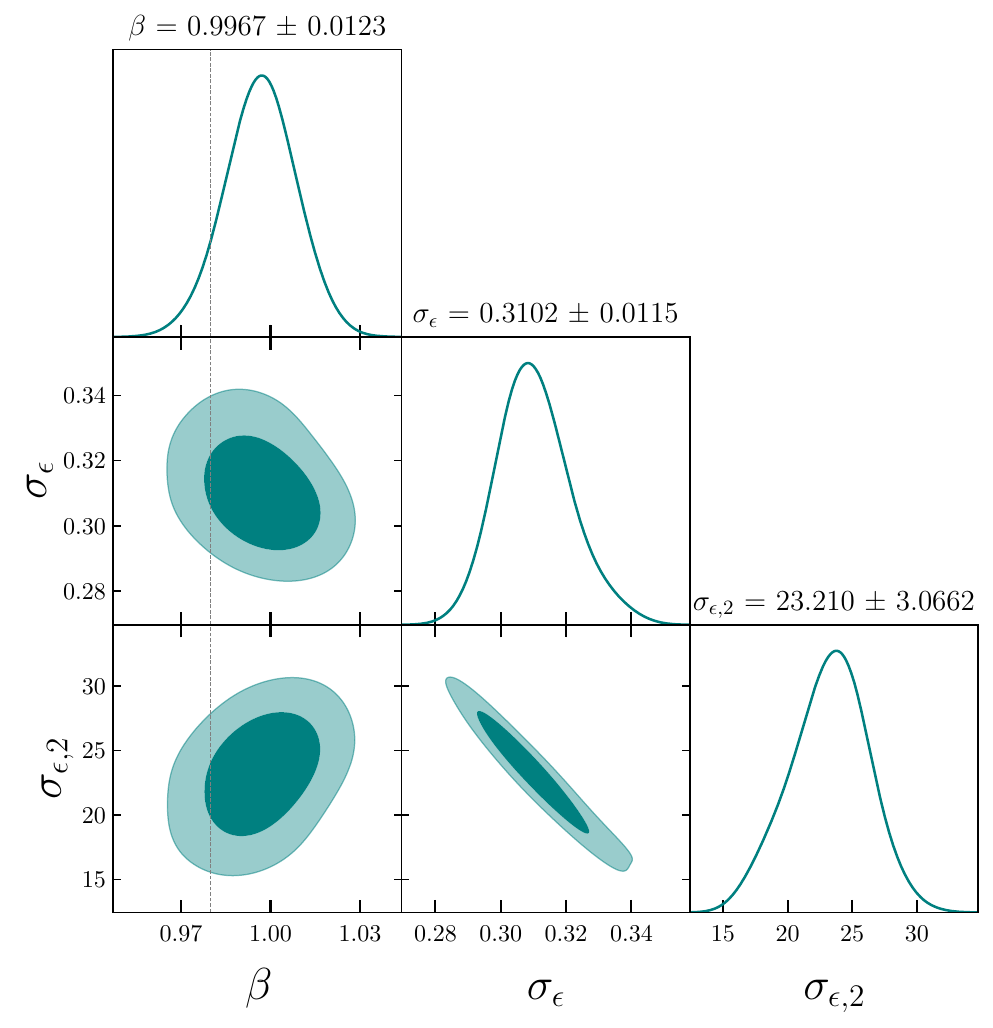}
  
    \caption{Lagrangian bias, $\Lambda = 0.15 \invMpc$.} 
  \end{subfigure}
  \begin{subfigure}[b]{0.49\textwidth}
    \centering
\includegraphics[width=\linewidth]{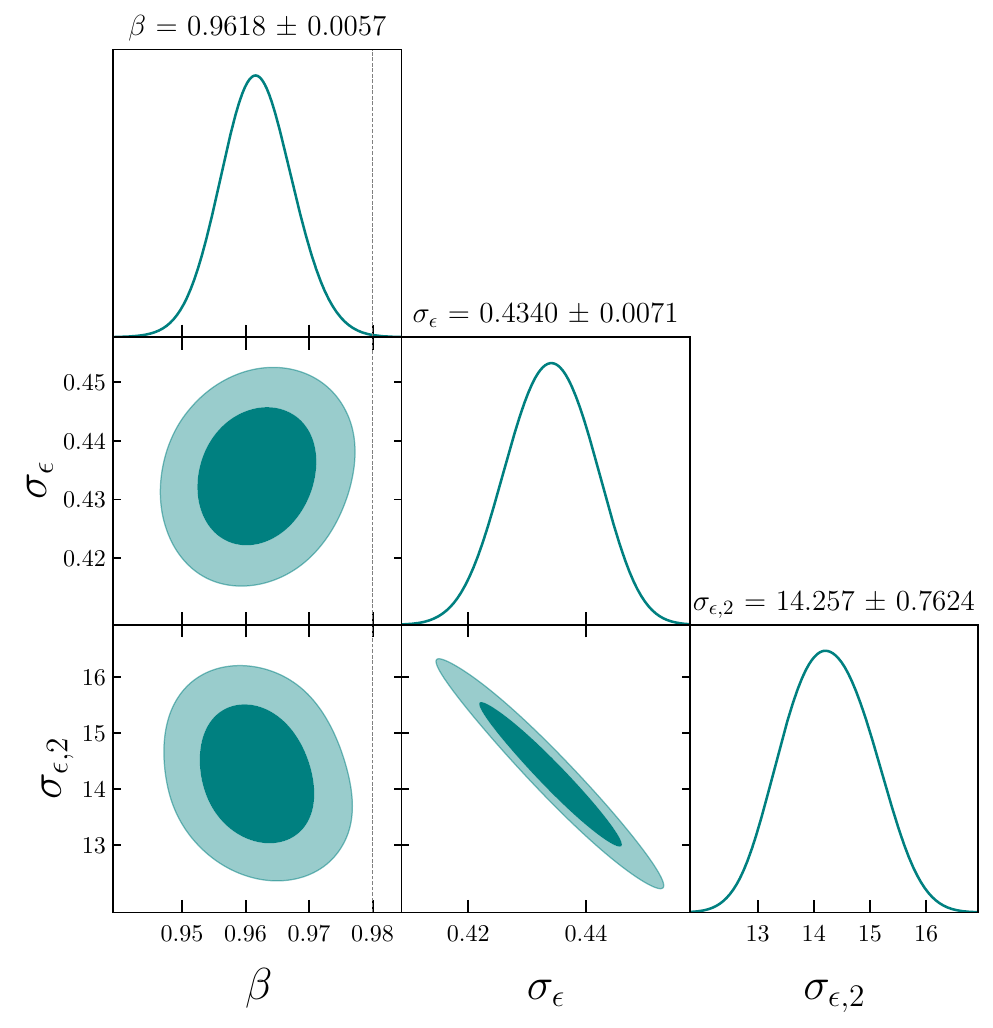}
   
    \caption{Lagrangian bias, $\Lambda = 0.2 \invMpc$.} 
  \end{subfigure} 
  \caption{Parameter posteriors for the inference performed on Mock B. Top two panels (orange color) correspond to Eulerian bias, while the lower two panels (blue) correspond to Lagrangian bias. The left panels show results for $\Lambda=0.15\invMpc$, while the right panels those for $\Lambda=0.2\invMpc$. The intermediate cutoff $\Lambda=0.18 \invMpc$ behaves similarly, and is not shown here for brevity.}
  \label{fig: Mock_B_corners}
\end{figure}

\begin{figure}
     \centering
     \begin{subfigure}[b]{0.49\textwidth}
         \centering
         \includegraphics[width=\linewidth]{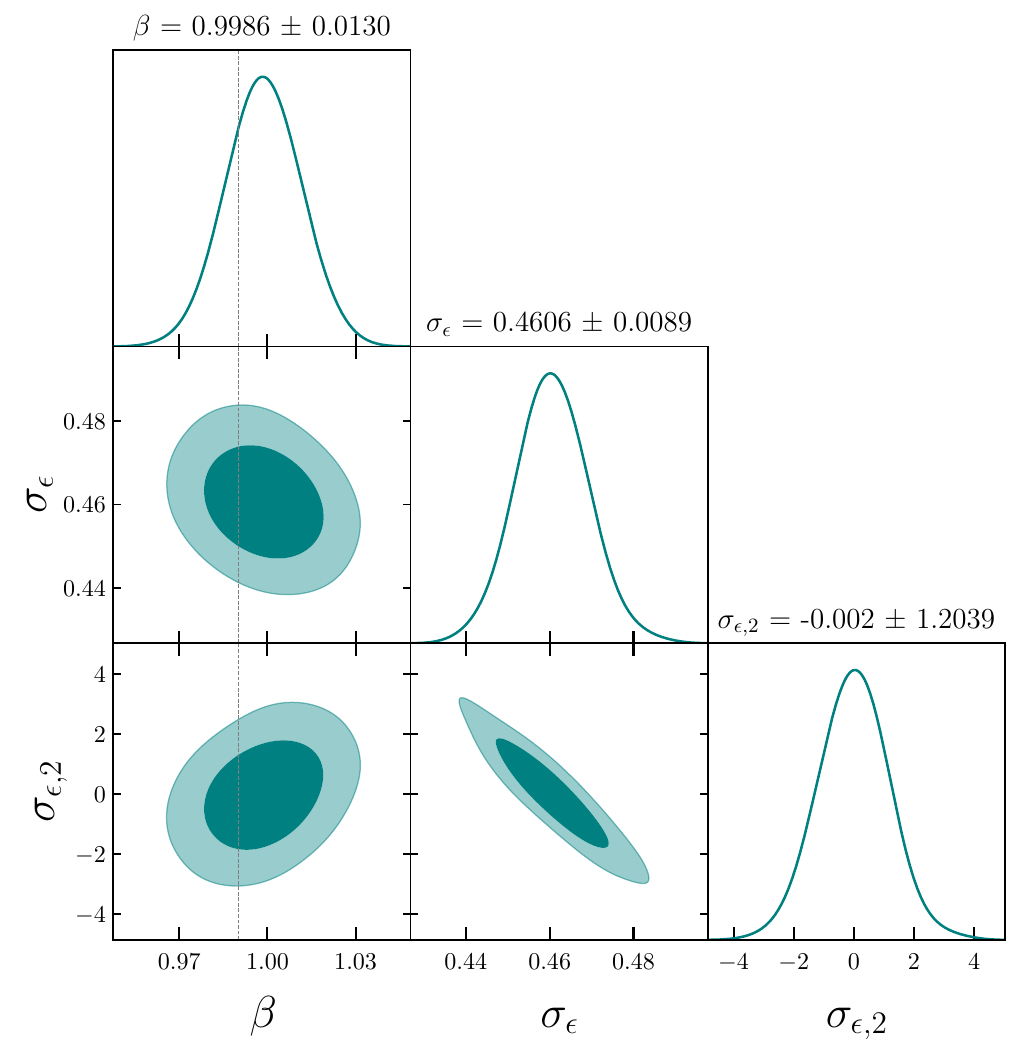}
    \caption{$\Lambda = 0.15 \invMpc$.} 
         \label{fig:y equals x}
     \end{subfigure}
     \hfill
     \begin{subfigure}[b]{0.49\textwidth}
         \centering
         \includegraphics[width=\linewidth]{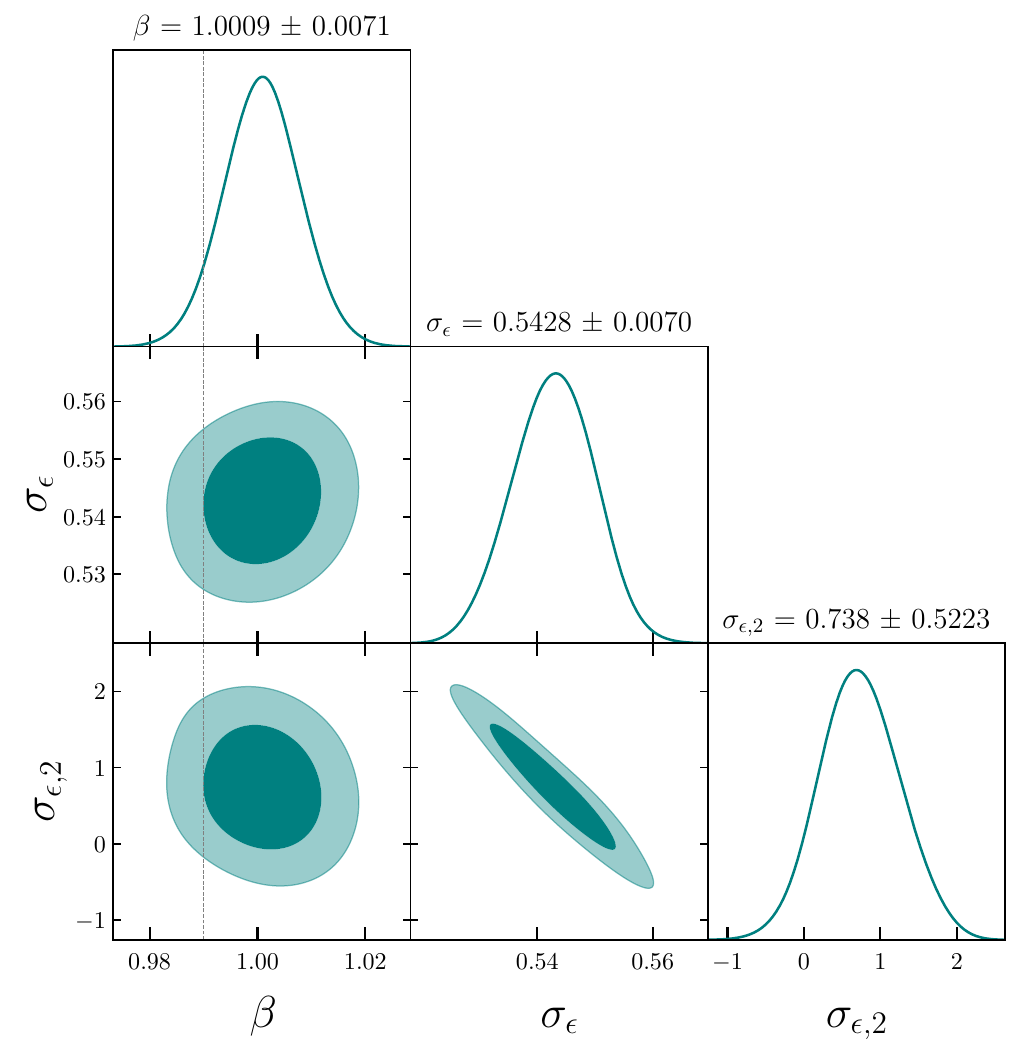}
    \caption{$\Lambda = 0.18 \invMpc$.}
     \end{subfigure}
     
     \begin{subfigure}[b]{0.49\textwidth}
         \centering
        \includegraphics[width=\linewidth]{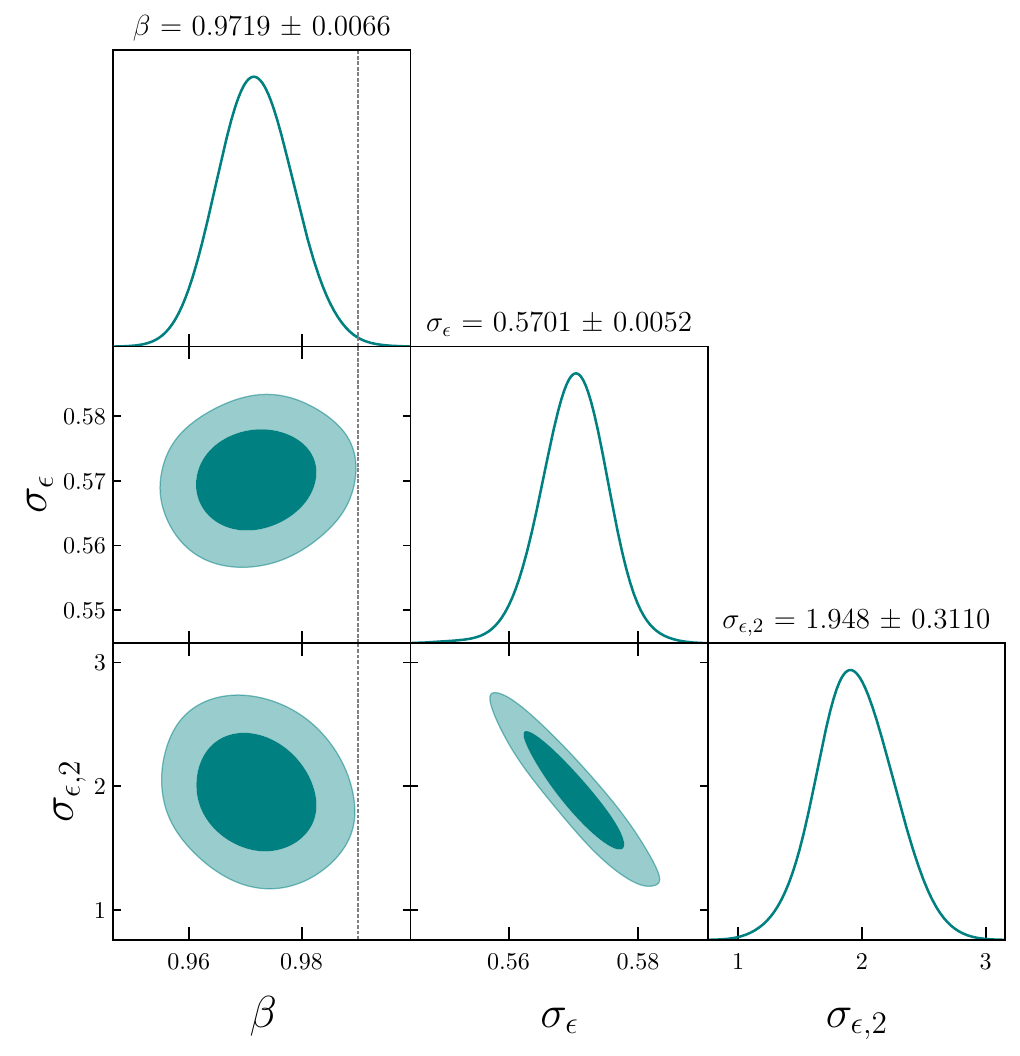}
  
    \caption{$\Lambda = 0.2 \invMpc$.}
         \label{fig:five over x}
     \end{subfigure}
        
  \caption{Parameter posteriors for the inference performed on Mock A.}\label{fig: Mock_A_corners}
\end{figure}

Below we present the corner plots showing posterior distribution of parameters sampled for both mocks and cutoffs. Results for Mock B are summarized in Fig. \ref{fig: Mock_B_corners}, where again we denote Lagrangian bias results in blue and Eulerian in orange.  
We notice the same trend of increasing noise as we did in Sec. \ref{Sec: results_field}.
In Fig. \ref{fig: Mock_A_corners}, we present the results for Mock A. The effective noise also grows in the case of Mock A, but in this case it is $\sigma_{\epsilon}$ that grows more significantly, rather than $\sigma_{\epsilon,2}$.

\subsection{Parameter posteriors: non-marginalized likelihood}
\begin{figure}[t]
    \centering
\includegraphics[width=0.6\textwidth]{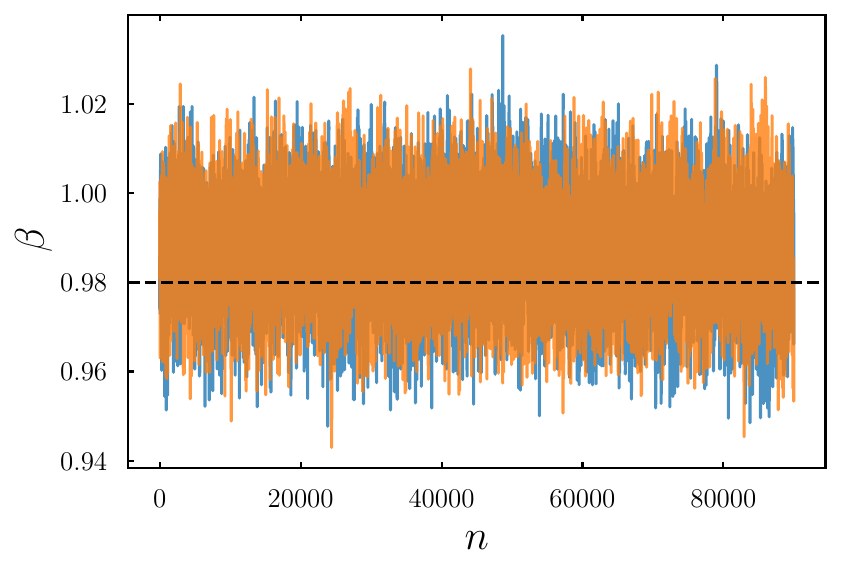}
    \caption{Trace plot for parameter $\beta$ in chains generated using the non-marginalized likelihood. Inference was performed at $\Lambda = 0.15 \invMpc$ for Mock B.}
    \label{fig: trace four B}
\end{figure}

In this section, we present the results of the inference obtained using a likelihood that was not marginalized over the bias parameters. For this analysis, uniform priors were applied to the bias parameters. Specifically, a uniform prior of $\mathcal{U}(0.01, 10)$ was used for $b_{\delta}$, while all other bias parameters were assigned a uniform prior of $\mathcal{U}(-30, 30)$. For the inference process, two chains per sample were initiated, each starting from random initial conditions $\hat s$ and a different initial value of $\beta$. These chains converged quickly to a similar $\beta$ value, as illustrated in Figure \ref{fig: trace four B}, which displays the trace plot of $\beta$ for Mock B sampled with Eulerian bias at $\Lambda = 0.15 \invMpc$. The chains were run until 100 effective samples were obtained. 

\begin{figure}[h!]
    \centering
\includegraphics[width=0.9\textwidth]{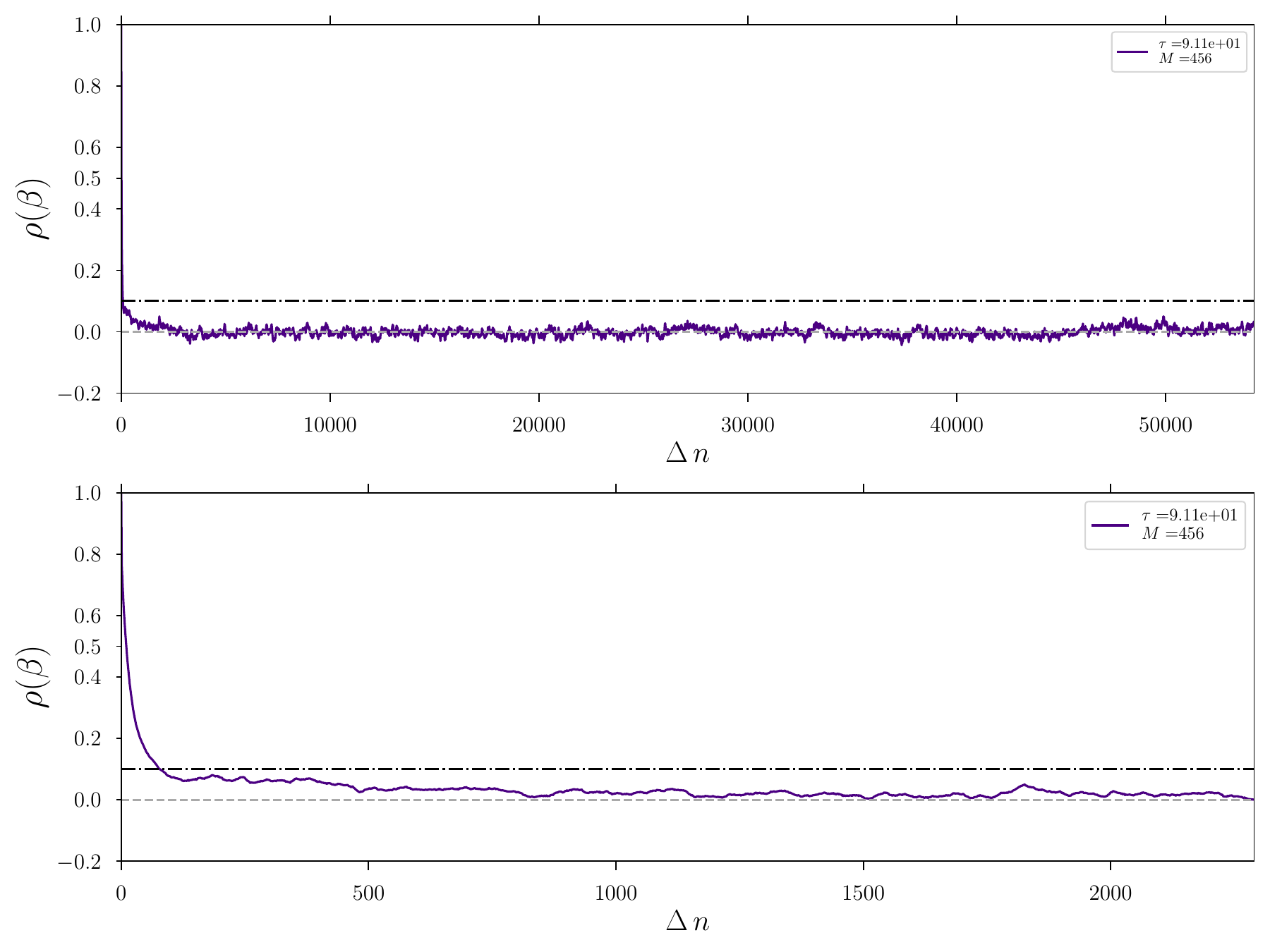}
    \caption{The normalized auto-correlation function for parameter $\beta$ inferred at $\Lambda = 0.15 \invMpc$ using the non-marginalized likelihood. We also show the correlation length value $\tau$ together with the maximum separation $M$ between the samples considered. The lower panel zooms in on the first 3000 samples.}
    \label{fig: corr B}
\end{figure}

In Fig. \ref{fig: corr B} we show the auto-correlation plot for one of the chains. We see that the correlation length for $\beta$ is very short even in the case when the non-marginalized likelihood is used. In fact, the correlation length is shorter than for the marginalized likelihood in this case (Tab.~\ref{T: T3}), in contrast to the findings for $\sigma_8$ in \cite{Kosti__2023}. This finding is clearly worth further investigation; note that the parameter posteriors show no significant correlations of $\beta$ with any other parameter.

\begin{figure}[t]
    \centering
\includegraphics[width=\textwidth]{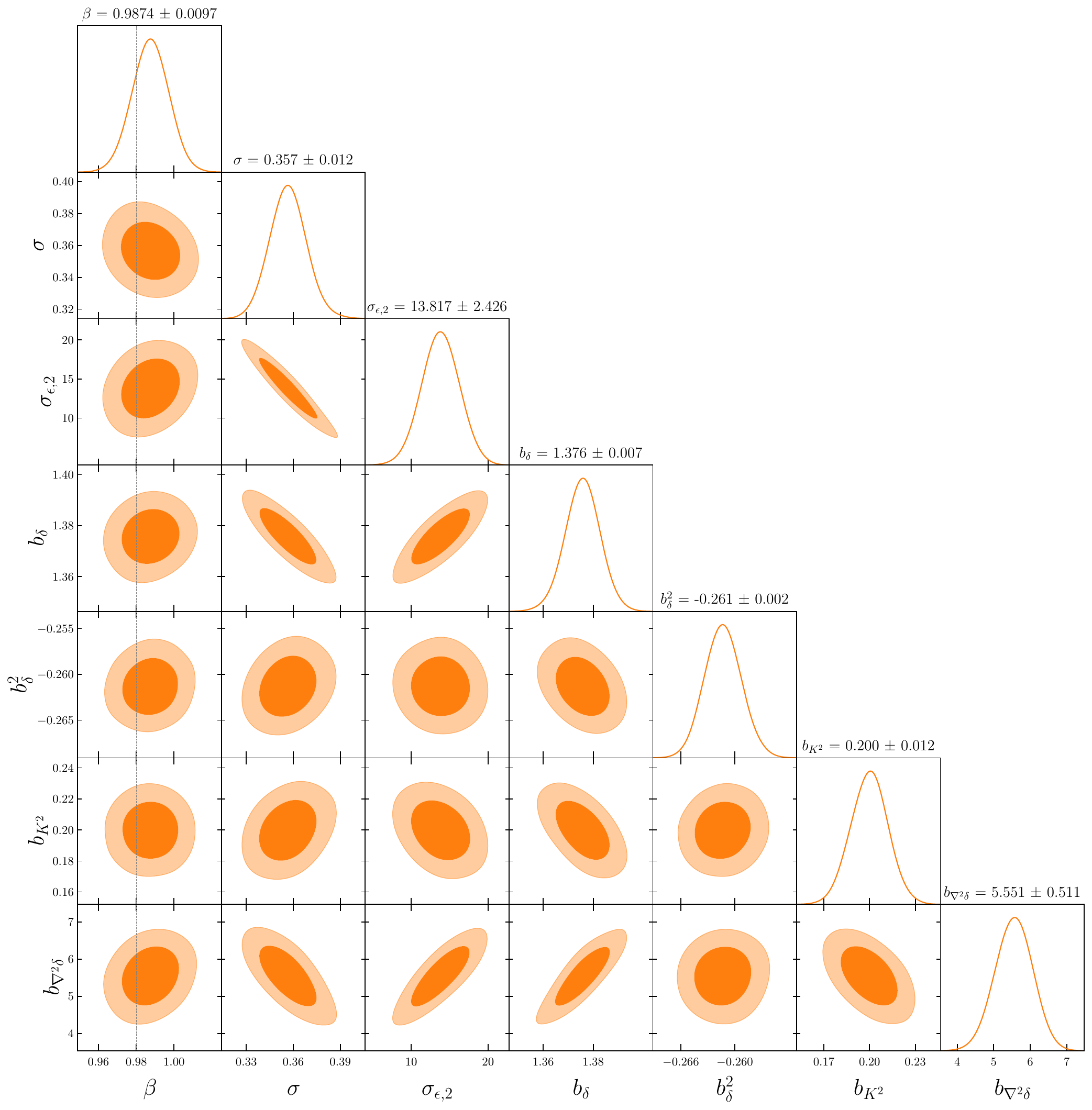}
    \caption{Results of the joint FreeIC inference on Mock B using non-marginalized likelihood and Eulerian bias. We show the posterior for all noise parameters, bias parameters and $\beta$. Inference was performed at $\Lambda = 0.15 \invMpc$.}
    \label{fig: Mock B Fourier}
\end{figure}

Finally, Figure \ref{fig: Mock B Fourier} displays the corner plots for all sampled parameters, including the mean value and the size of the error bars for each parameter. Comparing the size of the error bars for $\sigma(\beta)$ in this plot with those in Figure \ref{fig: marg_res} reveals that the error bar size for $\beta$ does not depend on whether bias parameters are sampled or marginalized analytically, as expected. Furthermore, the values of $\sigma_{\epsilon}$ and $\sigma_{\epsilon,2}$ are consistent with those obtained using the marginalized likelihood. 

The inferred values of the bias parameters differ from those used to generate the mock data, due to the mismatch in $\Lambda$ between the generation and sampling of the mock data \cite{2024JCAP...01..031R}. That is, the bias parameters and stochastic parameters are expected to run with $\Lambda$, while the inferred cosmological parameter, in this case $\beta$, should be consistent with the ground truth value.

\begin{figure}[t]
    \centering
\includegraphics[width=\textwidth]{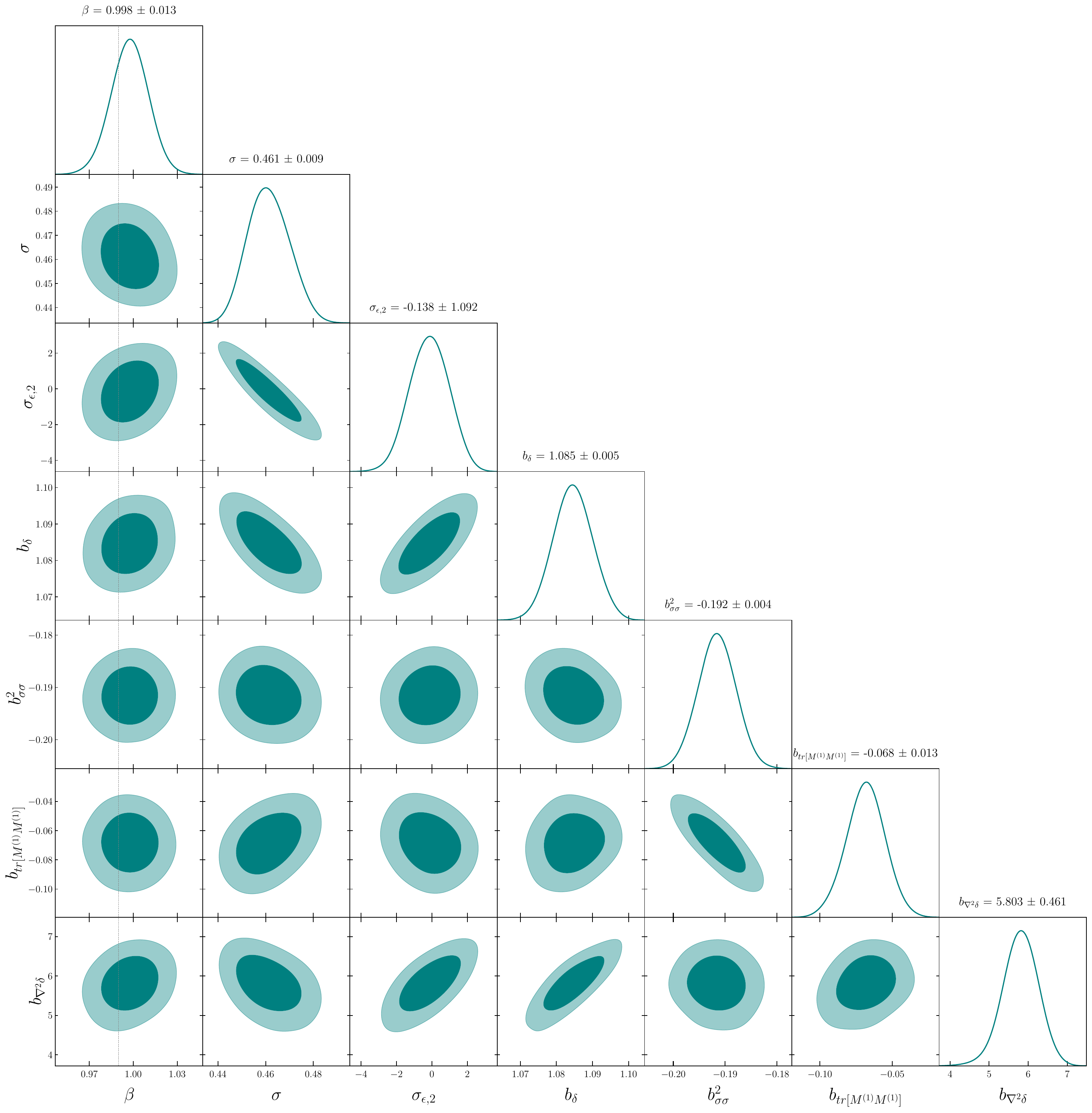}
    \caption{Results of the joint FreeIC inference on Mock A using non-marginalized likelihood and Lagrangian bias. We show the posterior for all noise parameters, bias parameters and $\beta$. Inference was performed at $\Lambda = 0.15 \invMpc$.}
    \label{fig: Mock A Fourier}
\end{figure}

The same analysis using the non-marginalized likelihood has been applied to Mock A sampled with the Lagrangian bias. The results of this analysis are summarized in Fig. \ref{fig: Mock A Fourier}. Similarly as in the case of mock B, we find that the error bar $\sigma(\beta)$ is the same as when bias parameters are marginalized analytically. 

% FS: --> Sec 5, and renamed
%\input{./Sections/App_2_Test_1LPT1D.tex}

\bibliographystyle{unsrt.bst}

\end{document}